\documentclass[12pt]{article}

\usepackage{bbm}
\usepackage{epsfig}
\usepackage{array}
\usepackage{float}
\usepackage{dsfont}
\usepackage{amstext}
\usepackage{rotating}
\usepackage{a4}
\usepackage{a4wide}
\usepackage{cite}

\def\ra{\rightarrow}
\def\be{\begin{equation}}
\def\ee{\end{equation}}
\def\gs{\mathrel{
   \rlap{\raise 0.511ex \hbox{$>$}}{\lower 0.511ex \hbox{$\sim$}}}}
\def\ls{\mathrel{
   \rlap{\raise 0.511ex \hbox{$<$}}{\lower 0.511ex \hbox{$\sim$}}}}

\newcommand{\onbb}{neutrinoless double beta decay }
\newcommand{\ba}{\begin{array}{c}}
\newcommand{\baz}{\begin{array}{cc}}
\newcommand{\bad}{\begin{array}{ccc}}
\newcommand{\bav}{\begin{array}{cccc}}
\newcommand{\baf}{\begin{array}{ccccc}}
\newcommand{\bea}{\begin{equation} \begin{array}{c}}
\newcommand{\eea}{ \end{array} \end{equation}}
\newcommand{\ea}{\end{array}}
\newcommand{\D}{\displaystyle}
\newcommand{\dms}{\mbox{$\Delta m^2_{\odot}$}}
\newcommand{\dma}{\mbox{$\Delta m^2_{\rm A}$}}
\newcommand{\meff}{\mbox{$\langle m \rangle$}}

\begin{document}

\title{
\hfill {\small HRI-P-08-07-003}
\vskip -0.3cm
\hfill {\small arXiv: 0807.4289 [hep-ph]}
\vskip 0.4cm
\Large \bf
Phenomenological consequences of four zero neutrino Yukawa textures}
\author{
Sandhya Choubey$^a$
~~,~~Werner Rodejohann$^b$
~~,~~Probir Roy$^{c*}$
\\\\
{\normalsize \it$^a$Harish--Chandra Research Institute,}\\
{\normalsize \it Chhatnag Road, Jhunsi, Allahabad 211019, India }\\ \\ 
{\normalsize \it$^b$Max--Planck--Institut f\"ur Kernphysik,}\\
{\normalsize \it  Postfach 103980, D--69029 Heidelberg, Germany} \\ \\
{\normalsize \it${^*}$DAE Raja Ramanna Fellow}\\
{\normalsize \it$^c$Saha Institute of Nuclear Physics,}\\
{\normalsize \it  Block AF, Sector 1, Kolkata 700 064, India}
}
\date{}
\maketitle
\thispagestyle{empty}
\vspace{-0.8cm}
\begin{abstract}
\noindent

For type I seesaw and in the basis where the charged lepton and 
heavy right-handed neutrino mass matrices are real and diagonal, 
four has been shown to be the maximum number of zeros 
allowed in the neutrino Yukawa coupling matrix $Y_\nu$. 
These four zero textures have been classified into 
two distinct categories. We investigate certain phenomenological 
consequences of these textures within a supersymmetric framework. 
This is done by using conditions implied on elements of the 
neutrino Majorana mass matrix for textures of each 
category in $Y_\nu$. These conditions turn out to be stable 
under radiative corrections. Including the effective 
mass, which appears 
in neutrinoless double beta decay, along with the usual neutrino masses, 
mixing angles and phases, it is shown analytically and through scatter 
plots how restricted regions in the seesaw parameter space 
are selected by these conditions. We also make 
consequential statements on the yet unobserved radiative 
lepton flavor violating decays such as $\mu \ra e \gamma$. All these 
decay amplitudes are proportional to the moduli of entries of the 
neutrino Majorana mass matrix. We also show under which conditions 
the low energy CP violation, showing up in neutrino oscillations, 
is directly linked to the CP violation required for producing 
successful flavor dependent and flavor independent 
lepton asymmetries during leptogenesis. 

\end{abstract}

\newpage

\section{\label{sec:intro}Introduction}

Neutrino mixing, leptogenesis and (in a supersymmetric framework) 
radiative lepton flavor violating decays 
$(\ell_i \rightarrow \ell_j  \gamma)$, $\ell$ being a charged lepton 
and $i,j$ being generation indices,  
have generally been acknowledged \cite{seesaw} as important tools 
to constrain parts of the seesaw \cite{I} parameter space. 
Any study of these tools gets considerably facilitated by
the assumption of texture zeros being present in the Yukawa coupling matrix
$Y_\nu = m_D /v_u$ \cite{FGM,IR,KT,6authors}. 
Here $v_u$ is the vev of the up-type Higgs and $m_D$
is the Dirac neutrino mass matrix. 
Texture zeros also help in relating
\cite{FGM,IR,KT,6authors} CP violation at low energies to that required 
for leptogenesis. 
Given the observed pattern of neutrino mixing and assuming no 
neutrino to be exactly massless, four is now known \cite{4zero} to 
be the maximum number of zeros allowed in $m_D$ within the 
type I seesaw framework. More zero entries in $Y_\nu$  
lead \cite{4zero} to at least one completely unmixed neutrino. 
This statement is made, of course, in the standard weak basis 
where the right-handed neutrino and charged lepton mass matrices, 
$M_R$ and $m_\ell$ respectively, are real and diagonal.  All such 
allowed four zero textures in $m_D$ have been 
completely classified \cite{4zero}. 
Our aim in this work is to study the implications of these allowed 
and completely classified four zero textures for radiative 
lepton flavor violating decays as well as for leptogenesis. 
We shall also make observations on related aspects of 
neutrino mixing and neutrinoless double beta decay. \\ 

In type I seesaw \cite{I} the low energy neutrino mass matrix in family
space obeys the `matching condition'
\be
m_\nu = -m_D\,M_R^{-1}\,m_D^T = U\,m_\nu^{\rm diag}\,U^T
\, ,
\label{eq:seesaw}
\ee
where $U$ is the PMNS matrix, parameterizable in terms of three angles
$\theta_{12}, \theta_{23}, \theta_{13}$ and three phases $\alpha, \beta,  
\delta$. Thus, 
\begin{equation} 
\label{eq:Upara}
 U =  \left(
 \begin{array}{ccc}
 c_{12} \, c_{13} & s_{12}\, c_{13} & s_{13}\, e^{-i \delta}\\
 -c_{23}\, s_{12}-s_{23}\, s_{13}\, c_{12}\, e^{i \delta} &
 c_{23}\, c_{12}-s_{23}\, s_{13}\, s_{12}\, 
e^{i \delta} & s_{23}\, c_{13}\\
 s_{23}\, s_{12}-\, c_{23}\, s_{13}\, c_{12}\, e^{i \delta} &
 -s_{23}\, c_{12}-c_{23}\, s_{13}\, s_{12}\, 
e^{i \delta} & c_{23}\, c_{13}
 \end{array}
 \right) P \, ,
\end{equation}
where $c_{ij} = \cos \theta_{ij}$, $s_{ij} = \sin \theta_{ij}$,  
$\delta$ is the Dirac-type CP-violating phase 
and the Majorana phases 
$\alpha$ and $\beta$ are contained in the matrix 
$P = {\rm diag}(1, e^{i \alpha}, e^{i (\beta + \delta)})$. 
Leptonic CP violation, that can show up in neutrino oscillation 
experiments, can be described through a rephasing (Jarlskog) 
invariant quantity given by \cite{6authors} 
\bea \label{eq:jcp0}
J_{\rm CP} = {\rm Im} 
\left\{ U_{e1} \, U_{\mu 2} \, U_{e 2}^\ast \, U_{\mu 1}^\ast \right\} 
= -\frac{\D {\rm Im} \left\{ h_{12} \, h_{23} \, h_{31} \right\} }
{\D \Delta m^2_{21} \, \Delta m^2_{31} \, \Delta m^2_{32}~}\, , 
\mbox{ where } ~h = m_\nu \, m_\nu^\dagger\,. 
\eea
With the parameterization of Eq.~(\ref{eq:Upara}), one has  
$J_{\rm CP} = \frac{1}{8} \, \sin 2 \theta_{12}\,  
\sin 2 \theta_{23}\, \sin 2 \theta_{13}\, 
\cos\theta_{13}\, \sin\delta$. 

Whereas the presence or the value of any of the phases is currently unknown, 
the oscillation parameters are constrained as follows \cite{data} (see 
also \cite{datanew}): 
\begin{eqnarray} \label{eq:data} \nonumber 
\Delta m^2_{21} & = &  7.67 \,_{-0.21}^{+0.22} \, 
\left(_{-0.61}^{+0.67}\right) \times 10^{-5}~{\rm eV}^2 \,, 
\\ \nonumber 
    \Delta m^2_{31} & = & \left\{ \baz 
-2.37 \pm 0.15 \,\left(_{-0.46}^{+0.43}\right) 
\times 10^{-3}~{\rm eV}^2  & \text{(inverted ordering)} \,, \\ 
+2.46 \pm 0.15 \,\left(_{-0.42}^{+0.47}\right) \times 
10^{-3}~{\rm eV}^2  & \text{(normal ordering)} \,,
    \ea \right. \\ 
\sin^2 \theta_{12} & = & 0.32 \pm 0.02 
\,\left(_{-0.06}^{+0.08}\right) \,,  \\ \nonumber 
    \sin^2 \theta_{23} & = & 0.45 \,_{-0.06}^{+0.09} \, 
\left(_{-0.13}^{+0.19}\right) \,,  \\ \nonumber 
 \sin^2 \theta_{13} & = & 0.0 \,^{+0.019}_{-0.000} \, 
\left(^{+0.05}_{-0.00}\right) \,.
 \end{eqnarray}
The 1$\sigma$ (3$\sigma$) ranges around the best-fit values have been given 
above.\\

 As in Ref.~\cite{4zero}, we consider textures in $m_D$ 
in the basis in which both $m_\ell = {\rm diag}(m_e,m_\mu,m_\tau)$ and 
$M_R = {\rm diag}(M_1,M_2,M_3)$  are 
real and diagonal. All flavor mixing information is thus encoded in 
the Dirac mass matrix $m_D$. The latter can be written in the 
most general form as 
\be
m_D 
= \left( 
\bad 
a_1 \, e^{i \alpha_1} & a_2 \, e^{i \alpha_2} & 
a_3 \, e^{i \alpha_3} \\
b_1 \, e^{i \beta_1} & b_2 \, e^{i \beta_2} & 
b_3 \, e^{i \beta_3} \\
c_1 \, e^{i \gamma_1} & c_2 \, e^{i \gamma_2} & 
c_3 \, e^{i \gamma_3} 
\ea
\right)\,.
\ee
Here, for each entry, we have listed the real amplitude $(a_i,b_i,c_i)$ 
and the corresponding phase $(\alpha_i,\beta_i,\gamma_i)$ explicitly.
Of course, three of the phases (one per row) can be rephased away. 
This Dirac mass matrix can also 
be expressed as \cite{CI}
\be
m_D = i \, U \, \sqrt{m_\nu^{\rm diag}} \, R \, \sqrt{M_R} \, ,
\label{eq:CI}
\ee
where $R$ is a complex, orthogonal matrix. This 
Casas-Ibarra parametrization illustrates an important feature: even 
when the elements of $m_\nu$ and $M_R$ are known, 
there is still an infinite number of Dirac mass matrices leading 
to the observed low energy phenomenology. Other observables need 
to be used in order to break this degeneracy \cite{rev}.\\

A well-known strategy to distinguish between 
different models, leading to 
the same low energy neutrino data, is to make use of 
Lepton Flavor Violation (LFV) and leptogenesis. 
LFV in supersymmetric seesaw scenarios leads to loop-induced decays such as 
$\ell_i \ra \ell_j \gamma$, with flavor indices $i,j$ spanning 
$(1=e,\,2=\mu,\,3=\tau)$, with the constraint $i > j$.
In mSUGRA scenarios, with 
universal boundary conditions for scalar sparticle mass 
matrices, one obtains the one-loop relation \cite{LFV} 
\be
{\rm BR}(\ell_i \rightarrow \ell_j + \gamma) = 
({\rm const}) \, {\rm BR}(\ell_i \ra \ell_j 
\, \nu \overline{\nu}) \, 
|(m_D \, L \,m_D^\dagger)_{ij}|^2\, ,
\label{eq:branching}
\ee
where the diagonal matrix $L$ is defined as 
\be
L_{kl} = \ln \frac{M_X}{M_k} \, \delta_{kl},
\label{eq:L}
\ee
with $M_k$ being the mass of the $k^{\rm th}$ right-handed 
neutrino. The logarithmic coefficient in the RHS of Eq.~(\ref{eq:L}) 
takes into account the effect of renormalization group 
running from a high scale $M_X$ 
to the scale of the respective heavy neutrino masses. 
The normalization factor ${\rm BR}(\ell_i \ra \ell_j 
\, \nu \overline{\nu})$ in the definition of 
the branching ratios in Eq.~(\ref{eq:branching}) is
noteworthy. The relevant 
numbers here 
are ${\rm BR}(\tau \ra e \, \nu \overline{\nu}) = 0.178$ 
and 
${\rm BR}(\tau \ra \mu \, \nu \overline{\nu}) = 0.174$ \cite{PDG}, 
respectively. For our later numerical work, we will ignore the small 
difference between the two. We will also take 
${\rm BR} (\mu \ra e \, \nu \overline{\nu})$ to be unity. 
Current upper limits on the branching ratios for 
$\ell_i \ra \ell_j \gamma$ are as follows: 
BR$(\mu \ra e \gamma) \le 1.2 \times 10^{-11}$ \cite{mueg_lim}, 
${\rm BR}(\tau \ra e \gamma) \le 
1.1 \times 10^{-7}$ \cite{teg_lim} and 
${\rm BR}(\tau \ra \mu \gamma) \le 6.8 \times 10^{-8}$ 
\cite{tmg_lim}. One expects these bounds to improve 
by two to three orders of magnitude for BR$(\mu \ra e \gamma)$ 
\cite{meg_fut} and by one to two orders of magnitude 
for the other branching ratios \cite{BR_fut} in the foreseeable future. 
The unspecified constant in the RHS of Eq.~(\ref{eq:branching}) depends 
on certain supersymmetry breaking parameters of mSUGRA, specifically the 
universal scalar and gaugino masses and the universal 
trilinear scalar coupling as well as on $\tan \beta$. 
However, we are not interested here in the exact magnitude of the 
branching ratios. We shall instead study the vanishing of certain 
branching ratios which for $\ell_i \ra \ell_j + \gamma$ turn 
out to be proportional to the square of the $i,j$th 
element of the low energy mass matrix $m_\nu$. \\

In principle, the above analysis could be extended also
to other lepton flavor violating processes, such as $\mu$--$e$
conversion in nuclei \cite{conv}. Current experimental limits on this
process, however, are expected to be improved considerably only
much after stronger limits on
$\ell_i \ra \ell_j \gamma$ have been made available.
If the photon penguin contribution dominates
the LFV diagrams, as happens for the case under study,
a good estimate for the ratio of BR$(\mu \ra e \gamma)$ to
the rate of $\mu$--$e$ conversion is ${\cal O}(1/\alpha)$, where
$\alpha$ is the electromagnetic fine structure constant.
In particular, the rate of $\mu$--$e$ conversion is also proportional
to $(m_D \, L \, m_D^\dagger)_{12}$. Hence, if in one of the
scenarios to be discussed BR($\mu \ra e \gamma$)
vanishes, $\mu$--$e$ conversion will be absent as well.
Note, moreover, that since only one conversion channel ($\mu \ra e$)
is experimentally accesible for the conversion process, no potentially
testable double ratios can be given. For these reasons, our focus here
is on the $\ell_i \ra \ell_j \gamma$
decays. \\

The other important aspect of seesaw phenomenology is 
leptogenesis. Of particular interest are the decay 
asymmetries \cite{leptog,D} that depend explicitly on the 
charged lepton flavor:
\bea 
\label{eq:epsIal}
\varepsilon_i^\alpha \D  
\equiv \frac{\D \Gamma (N_i \ra \phi \, \bar{l}_\alpha) -
\Gamma (N_i \ra \phi^\dagger \, l_\alpha)}
{\D  \sum\limits_\beta \Big[ \Gamma (N_i \ra \phi \, \bar{l}_\beta) + 
       \Gamma (N_i \ra \phi^\dagger \, l_\beta)\Big]}  \\ \D 
\,=\, \frac{1}{8 \pi \, v_u^2} 
\, \frac{1}{(m_D^\dagger \, m_D)_{ii}}  \, 
 \sum\limits_{j \neq i}  \, \left( 
{\cal I}_{ij}^\alpha \, f(M_j^2/M_i^2) + 
{\cal J}_{ij}^\alpha \, \frac{1}{1-M_j^2/M_i^2} 
\right)
\,\, ,
\eea
where
\bea 
\label{eq:calIJ}
{\cal I}_{ij}^\alpha = {\rm Im} \Big[ \big(m_D^\dagger \big)_{i \alpha} 
\, \big(m_D \big)_{\alpha j} \big(m_D^\dagger m_D \big)_{ij} \Big]~,~~
{\cal J}_{ij}^\alpha = {\rm Im} \Big[ 
%
%
(m_D^{\dagger})_{i \alpha} (m_D)_{\alpha j} (m_D^{\dagger} m_D)_{ji}
\Big] \,. 
\eea 
It is evident that 
${\cal I}_{ij}^\alpha = - {\cal I}_{ji}^\alpha$ and 
${\cal J}_{ij}^\alpha = - {\cal J}_{ji}^\alpha$. 
In the MSSM, the function $f(x)$ has the form \cite{covi}
\be
\D f(x) = 
\sqrt{x} \, \Big[ 
\frac{2}{1 - x} - \ln \Big( \frac{1+x}{x} \Big) 
 \Big] \,.
\ee

We have given quite general expressions above for the decay 
asymmetries 
that can accommodate any nontrivial role played by flavor effects 
\cite{flavor_flav}. Thus $\varepsilon_i^\alpha$ describes 
the decay of a heavy right-handed neutrino 
of mass $M_i$ into a charged lepton of flavor 
$\alpha = e, \mu, \tau$. 
When the lowest-mass heavy 
neutrino is much lighter than the other two, i.e.~$M_1 \ll M_{2,3}$, 
the lepton asymmetry is dominated by the decay 
of this lightest of the heavy neutrinos. In this case 
$f(M_j^2/M_1^2) 
\simeq - 3 \, M_1/M_j$. Moreover, only the first term 
proportional to 
${\cal I}_{1j}^\alpha$ in Eq.~(\ref{eq:epsIal}) is relevant 
then since the 
second term proportional to ${\cal J}_ {ij}^\alpha$ is 
suppressed by an additional power of $M_1/M_j$. 
Note furthermore that the 
second term in Eq.~(\ref{eq:epsIal}) vanishes when one 
sums over flavors to obtain the flavor independent decay asymmetry:
\bea 
\label{eq:epsI}
\varepsilon_i \D  
\,=\, \sum\limits_\alpha {\varepsilon_i^\alpha} 
\equiv 
\frac{
\D \sum\limits_\alpha 
\Big[ 
\Gamma (N_i \ra \phi \, \bar{l}_\alpha) - 
\Gamma (N_i \ra \phi^\dagger \, l_\alpha)  \Big]}
{\D \sum\limits_\beta \Big[
\Gamma (N_i \ra \phi \, \bar{l}_\beta) + 
       \Gamma (N_i \ra \phi^\dagger \, l_\beta)\Big]} \\ 
\,=\, \D 
\frac{1}{8 \pi \, v_u^2} \, 
\frac{1}{(m_D^\dagger \, m_D)_{ii}}  
\, \sum\limits_{j \neq i} 
{\rm Im} \,
\Big[ \big(m_D^\dagger \, m_D \big)^2_{i j}\Big] 
\, f(M_j^2/M_i^2) 
\\ \D 
= \frac{1}{8 \pi \, v_u^2} \, 
\frac{1}{(m_D^\dagger \, m_D)_{ii}}  \, 
 {\cal I}_{ij}
\, ,
\eea
where we have defined  
\be
{\cal I}_{ij} = \sum\limits_\alpha  {\cal I}_{ij}^\alpha\,. 
\ee 
We note here already that for all 72 four zero textures that we
study the ${\cal J}_{1j}^\alpha$ vanish.
The expressions given above for the decay asymmetries are valid for 
the MSSM. Their flavor structure is, however, 
identical to that of just the Standard Model.\\

Equally important 
in leptogenesis are effective mass parameters that are responsible 
for the wash-out. We assume that a single 
heavy neutrino of mass   
$M_1$ is relevant for leptogenesis. Then every decay asymmetry 
$\varepsilon_1^\alpha$ gets washed out by an effective mass 
\be
\tilde{m}_1^\alpha = 
\frac{(m_D^\dagger)_{1 \alpha}\, (m_D)_{\alpha 1}}{M_1}\,.
\ee
Moreover, the wash-out can be estimated by inserting this effective 
mass in the function \cite{flavor_flav1} 
\be
\eta(x) \simeq 
\left( 
\frac{8.25 \times 10^{-3}~{\rm eV}}{x} + 
\left(\frac{x}{2 \times 10^{-4}~{\rm eV}}\right)^{1.16} 
\right)^{-1}\,.
\ee
The summation of $\tilde{m}_1^\alpha$ over the flavor index $\alpha$ 
yields $\tilde{m}_1$, which is the relevant parameter for the 
wash-out of $\varepsilon_1$. The final baryon asymmetry is 
\cite{flavor_flav0,flavor_flav,flavor_flav1} 
\be
Y_B \simeq \left\{ 
\baz
-0.01 \, \varepsilon_1 \, \eta(\tilde{m}_1) 
& \mbox{one-flavor}\, ,\\[0.2cm]
-0.003 
\left( 
(\varepsilon_1^e + \varepsilon_1^\mu) \, \eta\left(\frac{417}{589} 
(\tilde{m}_1^e + \tilde{m}_1^\mu) \right) + 
\varepsilon_1^\tau \, \eta \left(\frac{390}{589} 
\tilde{m}_1^\tau \right) \right) 
& \mbox{two-flavor}\, ,\\[0.2cm]
-0.003 
\left( 
\varepsilon_1^e \, \eta \left(\frac{151}{179} 
\, \tilde{m}_1^e  \right) + 
\varepsilon_1^\mu \, \eta \left(\frac{344}{537} \, \tilde{m}_1^\mu \right) + 
\varepsilon_1^\tau \, \eta \left(\frac{344}{537} \, 
\tilde{m}_1^\tau \right) \right) & \mbox{three-flavor}\,.
\ea \right.
\ee 
Here 
we have given separate expressions 
for one-, two- and three-flavored 
leptogenesis. The three-flavor case occurs for 
$M_1 \, (1 + \tan^2 \beta) \le 10^9$ GeV, 
the one-flavor case for 
$M_1 \, (1 + \tan^2 \beta) \ge 10^{12}$ GeV, 
and the two-flavor case (with the tau-flavor decoupling first 
and the sum of electron- and muon-flavors, which act indistinguishably) 
applies in between.\\ 

LFV and leptogenesis provide means of breaking degeneracies in the seesaw 
parameter space. This comes about since their dependence on the 
seesaw parameters is complementary to that of $m_\nu$. 
A related issue is the question of circumstances 
under which there is a 
connection between high and low energy CP violation, i.e., between the 
phases responsible for leptogenesis and the ones responsible for 
CP asymmetries in neutrino oscillations. 
Inasmuch as texture zeros simplify this process, the motivation behind 
the present study is to phenomenologically investigate the extent of this 
degeneracy breaking for all allowed Dirac mass matrices with 
four zero textures classified in \cite{4zero}. The rest of the paper is 
organized as follows: in Section \ref{sec:textures} 
the two categories, (i) and (ii), of the four zero textures in 
$m_D$ are recapitulated and the radiative 
stability of the corresponding conditions on $m_\nu$ is emphasized. In 
Section  \ref{sec:phenomenology} we discuss the phenomenology of 
these textures, focusing on lepton mixing and the ratio of ratios in 
radiative LFV decays as well as on leptogenesis, wash-out factors and the 
basis invariant Jarlskog CP-violating parameter $J_{\rm CP}$; 
subsections \ref{sec:1} and \ref{sec:2} cover categories (i) 
and (ii) respectively. The final Section \ref{sec:concl} 
contains a summary of our 
results and conclusions derived therefrom.

\section{\label{sec:textures}The Two Categories of Four Zero Textures}

It will be helpful to 
provide first a summary of the classification of the 
four zero textures. As enumerated in 
Ref.~\cite{4zero}, there are 72 allowed textures of this kind:
\begin{itemize}
\item[(i)] 54 textures in which two rows of $m_D$ are orthogonal 
element by element. 
They can be further divided into three subclasses containing 
18 matrices each:
\begin{itemize}
\item[(ia)] 18 textures in which the first and second row are 
orthogonal element by element, resulting in 
\be \label{eq:12zero}
(m_\nu)_{12} = (m_\nu)_{21} = 0 \, ,
\ee
i.e., the vanishing of the off-diagonal $12$ (or $e\mu$) entry of the 
effective neutrino Majorana mass matrix; 

\item[(ib)] 18 textures in which the first and third row are 
orthogonal element by element, resulting in 
the vanishing off-diagonal element condition 
\be \label{eq:13zero}
(m_\nu)_{13} = (m_\nu)_{31} = 0 \, ,
\ee
i.e., the vanishing of the off-diagonal $13$ (or $e\tau$) entry of the  
effective neutrino Majorana mass matrix;
 
\item[(ic)] 18 textures in which the second and third row are 
orthogonal element by element, resulting in 
\be \label{eq:23zero}
(m_\nu)_{23} = (m_\nu)_{32} = 0 \, ,
\ee
i.e., the vanishing of the off-diagonal 
$23$ (or $\mu\tau$) entry of the 
effective neutrino Majorana mass matrix;  
\end{itemize}

\item[(ii)] 18 textures in which two columns of $m_D$ 
are orthogonal element by element. 
They can be further divided into three subclasses containing 
6 matrices each:
\begin{itemize}
\item[(iia)] 6 textures with two zeros in the first row, 
resulting in the vanishing sub-determinant conditions 
\be \label{eq:2a}
|(m_\nu)_{11} \, (m_\nu)_{23}| - |(m_\nu)_{21} \, (m_\nu)_{13}| = 
{\rm arg} \left\{ (m_\nu)_{11} \, (m_\nu)_{23} \, 
(m_\nu)_{21}^\ast \, (m_\nu)_{13}^\ast \right\} = 0~;
\ee
\item[(iib)] 6 textures 
with two zeros in the second row, 
resulting in the vanishing sub-determinant conditions 
\be \label{eq:2b}
|(m_\nu)_{22} \, (m_\nu)_{13}| - |(m_\nu)_{12} \, (m_\nu)_{23}| = 
{\rm arg} \left\{ (m_\nu)_{22} \, (m_\nu)_{13} \, 
(m_\nu)_{12}^\ast \, (m_\nu)_{23}^\ast \right\} = 0~;
\ee
\item[(iic)] 6 textures with two zeros in the third row, 
resulting in the vanishing sub-determinant conditions 
\be \label{eq:2c}
|(m_\nu)_{33} \, (m_\nu)_{12}| - |(m_\nu)_{13} \, (m_\nu)_{32}| = 
{\rm arg} \left\{ (m_\nu)_{33} \, (m_\nu)_{12} \, 
(m_\nu)_{13}^\ast \, (m_\nu)_{32}^\ast \right\} = 0~.
\ee
\end{itemize}
\end{itemize}

We now wish to comment on the question of the dependence of such results, as 
presented above, on the energy scale. In general, elements of the matrix 
$Y_\nu$ change with energy in a coupled way due to radiative corrections 
leading to renormalization group (RG) running. Thus an element, which vanishes 
at low energies, can certainly develop a significant nonzero value at a very 
high energy. Our postulate is that four zeros are present in $Y_\nu$ at 
energies relevant to oscillation experiments. Within a reasonable accuracy, 
such can also be taken to be the case then for radiative LFV decays. This 
assumption is, however, generally not valid for leptogenesis which we take to 
operate at $M_1 \geq 10^9$ GeV. In a grand unified theory, of course, $Y_\nu$ 
would originate at $M_X \simeq 10^{16}$ GeV and would 
need to be evolved down to an energy scale below the $Z$-mass $m_Z$. 
If texture zeros are due to some (yet unknown) flavor symmetry at 
some high scale, one would need to assume that the 
said zeros are protected by the same symmetry during the RG running. If such 
is the case, our statements would continue to hold without modification. \\ 

Let us nevertheless point out a particularly interesting feature 
of the consequences for the neutrino Majorana mass matrix of 
the textures in $m_D$ under consideration. These conditions 
on $m_\nu$ are stable under RG running. 
If one performs the running from the high scale $M_X$ to the low scale 
$m_Z$, then for $m_\nu$ this can be taken into 
account by multiplying each matrix element 
$(m_\nu)_{ij}$ by a factor \cite{RGE}. The latter is given by 
 $(1 + \epsilon_i) \, (1 + \epsilon_j)$, 
where 
\[ 
\epsilon_i = c \, \frac{m_i^2}{16 \pi^2 \, v^2} \ln \frac{M_X}{m_Z}
\]
with $m_{3,2,1} = m_{\tau, \mu, e}$ being the charged lepton masses. 
The parameter $c$ is given by 3/2 in the SM and by 
$-(1 + \tan^2 \beta)$ in the MSSM. 
The multiplicative nature of this correction ensures that a 
vanishing element of $m_\nu$ stays vanishing. Thus the 
consequence for $m_\nu$ of every texture under category (i) is safe under 
RG running. It is, additionally, rather surprising that the 
corresponding consequences for $m_\nu$ from all textures
in category (ii) are also unharmed by the RG running from 
radiative corrections. 
For instance, consider category (iia) and perform the 
corrections in condition (\ref{eq:2a}). The multiplication 
of $(m_\nu)_{ij}$ with the factors $(1 + \epsilon_i) \, (1 + \epsilon_j)$
leads to 
\[
\ba \D 
|(m_\nu)_{11} \, (m_\nu)_{23}| 
- |(m_\nu)_{21} \, (m_\nu)_{13}| 
\rightarrow 
|(m_\nu)_{11}' \, (m_\nu)_{23}'| 
- |(m_\nu)_{21}' \, (m_\nu)_{13}'| \\ \D 
= |(m_\nu)_{11} \, (m_\nu)_{23}| \, (1 + \epsilon_1)^2 \, 
(1 + \epsilon_2) \, (1 + \epsilon_3) 
- |(m_\nu)_{21} \, (m_\nu)_{13}| \, 
(1 + \epsilon_1) \, (1 + \epsilon_2) \, 
(1 + \epsilon_1) \, (1 + \epsilon_3) \\ \D 
= (1 + \epsilon_1)^2 \, 
(1 + \epsilon_2) \, (1 + \epsilon_3) \, 
\left( |(m_\nu)_{11} \, (m_\nu)_{23}| 
- |(m_\nu)_{21} \, (m_\nu)_{13}| \right) \, ,
\ea
\]
so that condition (\ref{eq:2a}) remains unchanged. 
The result is identical for categories (iib) and (iic), the matrix indices 
$1,\,2,\,3$ appearing the same number of times on both sides of the 
conditions (\ref{eq:2b}) and (\ref{eq:2c}). This defines an interesting 
class of ``RG invariants''. 

\section{\label{sec:phenomenology}The Phenomenology of Four Zero Textures}

\subsection{\label{sec:1}Category (i)}

We first discuss some general issues concerning the 
phenomenology of the textures under consideration. 
To start with, take the subclass in category (i) in which rows 
$i$ and $j~(\neq i)$ of $m_D$ are orthogonal, element by element:
\be
(m_\nu)_{ij} = (m_\nu)_{ji} = 0 \,.
\ee
It also follows that for this subclass of four zero textures,
\be
(m_D \, K \, m_D^\dagger)_{ij} = 0\, , 
\ee
where $K$ is any diagonal matrix. The immediate implication 
is that the branching ratio for the decay 
$\ell_i \ra \ell_j \gamma$ is zero for these textures:  
\be \label{eq:lfvcat1}
{\rm BR}(\ell_i \rightarrow \ell_j \gamma) = 0\,.
\ee
Therefore all such textures of $m_D$ would be excluded by any future 
experimental observation of this decay 
mode\footnote{The requirement of a vanishing 
$(m_D \, m_D^\dagger)_{12}$ can 
lead via 2-loop effects to a lower limit 
on BR$(\mu \ra e \gamma)$, connected to the product of the 
branching ratios of $\tau \ra \mu \gamma$ and 
$\tau \ra e \gamma$ \cite{zero}. Obviously the 2-loop 
induced branching ratio is very small.}.\\

Another interesting 
feature is that, in the textures of category (i) and for an arbitrary
$j \neq i$, the relation 
\be 
|(m_D \, L \, m_D^\dagger)_{ij}| \propto |(m_\nu)_{ij}|~
\ee
holds. However, it is not possible to construct predictive 
ratios of branching ratios from this, unless the heavy neutrino 
masses are known. Consider as an example the texture 
\be \label{eq:ex1}
m_D = \left( 
\bad 
0 & a_2 & a_3 \, e^{i \alpha_3} \\ 
b_1 & 0 & 0 \\
c_1 & c_2 \, e^{i \gamma_2} & 0 \\
\ea
\right)\, ,
\ee
which belongs to category (ia) and 
leads to $(m_\nu)_{12} = {\rm BR}(\mu \ra e \gamma) = 0$. 
The non-zero branching ratios for the decays $\tau \ra e \gamma$ 
and $\tau \ra \mu \gamma$ are governed by 
\be 
\left|(m_D \, L \, m_D^\dagger)_{13}\right|^2 
= a_2^2 \, c_2^2 \, L_2^2 ~~\mbox{ and }~
\left|(m_D \, L \, m_D^\dagger)_{23}\right|^2 
= b_1^2 \, c_1^2 \, L_1^2\, ,
\ee
respectively. 
The low energy Majorana mass matrix is 
\be
m_\nu = -
\left( 
\bad \D 
\frac{a_2^2}{M_2} + \frac{a_3^2 \, e^{2 i \alpha_3}}{M_3} & 0 & \D 
\frac{a_2 \, c_2 \, e^{i \gamma_2}}{M_2} \\ \D 
\cdot & \D \frac{b_1^2}{M_1} & \D \frac{b_1 \, c_1}{M_1} \\ \D 
\cdot & \cdot & \D 
\frac{c_1^2}{M_1} + \frac{c_2^2 \, e^{2 i \gamma_2}}{M_2}
\ea
\right)\,.
\ee
It follows that 
\be
\frac{{\rm BR}(\tau \ra e \gamma)}{{\rm BR}(\tau \ra \mu \gamma)} 
= \frac{\left|(m_\nu)_{13}\right|^2}{\left|(m_\nu)_{23}\right|^2} 
\, \left(\frac{M_2}{M_1} \frac{L_2}{L_1} \right)^2\,.
\ee
Without any further 
information about the heavy neutrino masses, one is unable 
to predict this ratio. The same feature is 
valid for all textures in category (i). We will not work out here all 72 
possibilities. \\

{Leptogenesis -- either of the unflavored or of the 
flavor dependent variety -- is quite possible in general for such 
textures. 
As mentioned earlier, we work  under the 
assumption that $M_1 \ll M_{2,3}$ so that one needs to consider 
only the decay of 
$N_1$. Thus, among the parameters responsible for leptogenesis, 
$i$ is always 1. Generalization to the more general situation, 
including $M_{2,3}$ is, however, straightforward. 
It turns out that for the textures under study here 
all ${\cal J}_{1j}^\alpha$ vanish and we 
are left only with the ${\cal I}_{1j}^\alpha$. 
Specific textures in this subclass 
will of course have some of the ${\cal I}_{ij}^\alpha$ zero 
in case the appropriate elements of 
$m_D$ happen to lead to this.} In the example from 
Eq.~(\ref{eq:ex1}) the only non-zero decay asymmetry is 
\be \label{eq:ybexam1}
{\cal I}_{12}^\tau = c_1^2 \, c_2^2  \, \sin 2 \gamma_2 \,.
\ee  
It is interesting to ask under what circumstances the 
``leptogenesis phase'' $\gamma_2$ is responsible for low energy 
leptonic CP violation as well. Evaluating the invariant 
in Eq.~(\ref{eq:jcp0}), which describes CP violation in neutrino 
oscillations, gives 
\be \label{eq:jcpexam1}
\ba 
\Delta m_{21}^2 \, \Delta m_{31}^2 \, \Delta m_{32}^2 \, J_{\rm CP} 
= \D \frac{-a_2^2 \, b_1^2 \, c_1^2 \, c_2^2}{M_1^3 \, M_2^3 \, M_3} 
\,  \Big[ a_3^2\left(c_2^2 \, M_1 \, \sin 2 {\alpha_3} 
   +   \left(b_1^2 + c_1^2 \right) 
M_2 \, \sin 2(\alpha_3- \gamma_2) \right) 
\\ - \left(\left(b_1^2 + c_1^2\right) 
    a_2^2 + b_1^2 \, c_2^2\right) M_3 \, \sin 2  \gamma_2)\Big]\,.
\ea 
\ee 
It follows that the leptogenesis phase is related to the 
low energy Dirac phase when the conditions 
\be \label{eq:condexam1} 
\left(\left(b_1^2 + c_1^2 \right)  
    a_2^2 + b_1^2 \, c_2^2\right) M_3 
\gg a_3^2 \left| c_2^2 \, M_1 \, \sin 2 {\alpha_3} \right| 
~,~~ a_3^2 \left| \left(b_1^2 + c_1^2 \right)  M_2 
\, \sin 2(\alpha_3- \gamma_2)\right| 
\ee 
are fulfilled. Similar considerations can be made for all other 
textures in category (i). Note that in the basis that we are working, 
the Dirac mass matrix from the example in 
Eq.~(\ref{eq:ex1}) contains only two physical phases, which is one less  
than the number of low energy phases in $m_\nu$. This facilitates the 
connection of low and high energy CP violation. 
We stress here that 
{\it all} four zero textures contain two physical phases. For this reason 
all 72 candidates have -- in analogy to the conditions in 
Eq.~(\ref{eq:condexam1}) -- the possibility \cite{6authors}
that the low energy leptonic CP 
violating phases are the ones responsible for leptogenesis.\\

\begin{figure}[t]
\begin{center}
\epsfig{file=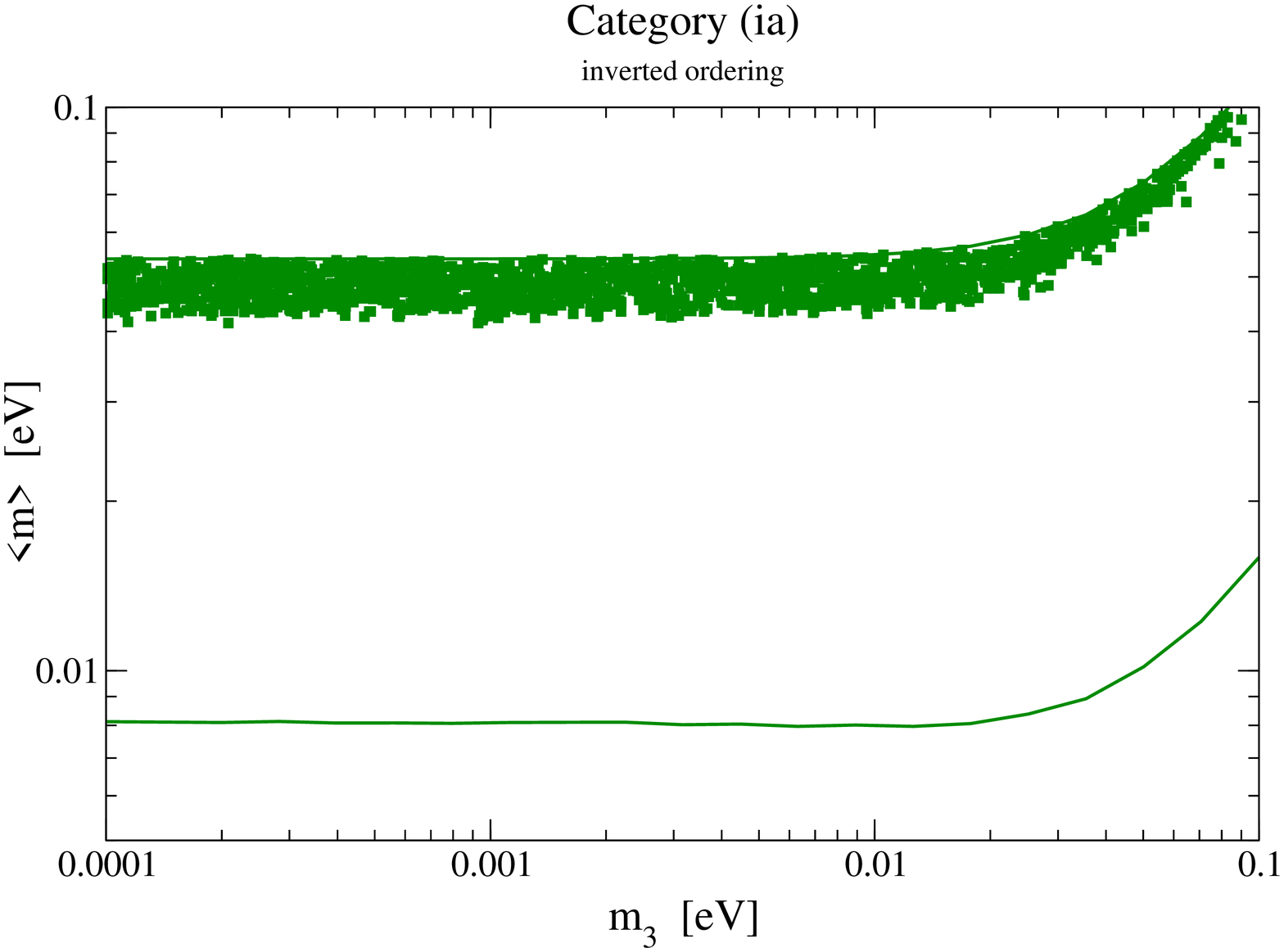,width=8cm,height=8cm}
\epsfig{file=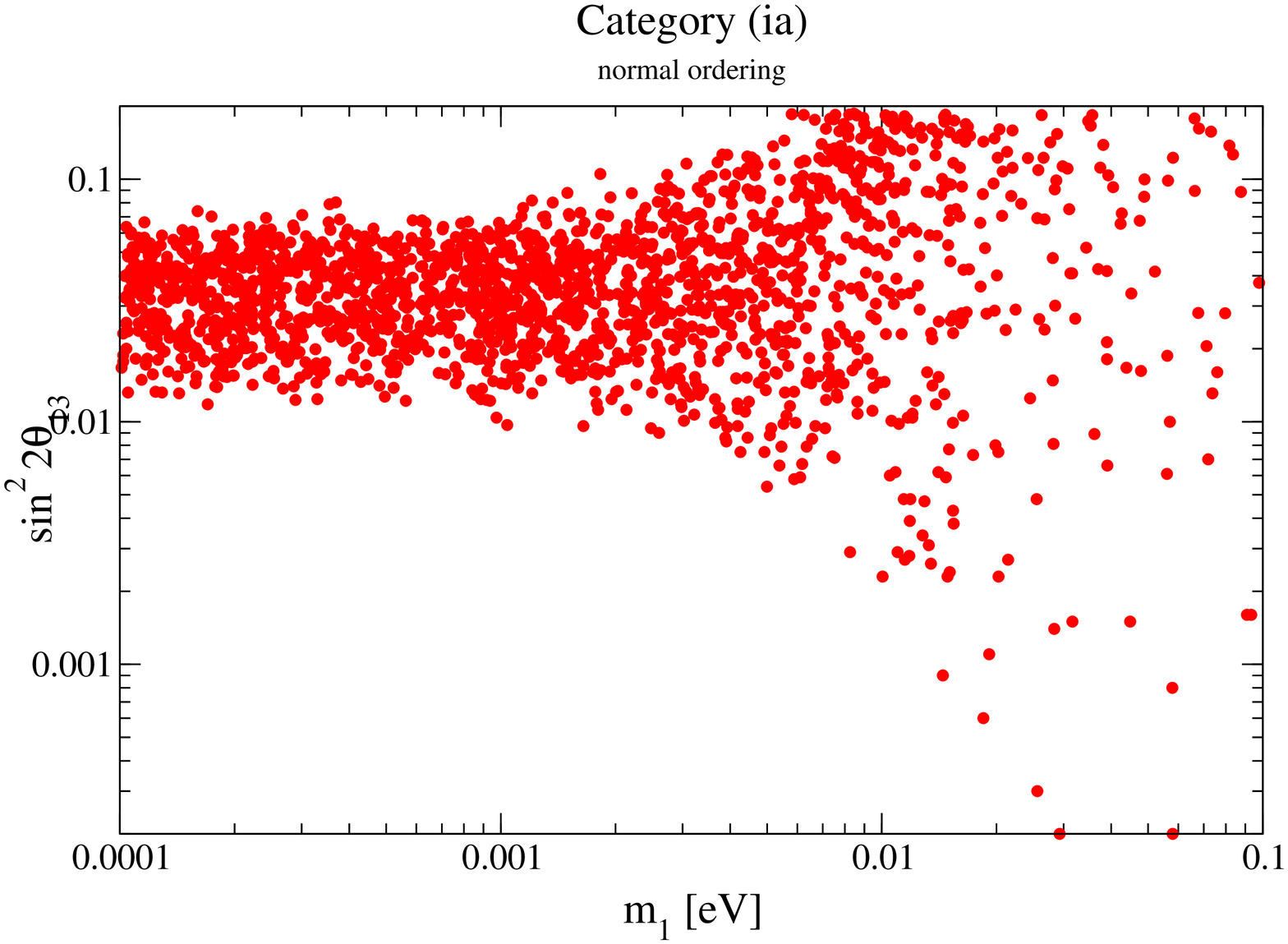,width=8cm,height=8cm}
\caption{\label{fig:1a}Category (ia) or $(m_\nu)_{12}=0$: 
scatter plots of the effective mass versus 
the smallest mass in 
case of an inverted mass ordering (left) and of 
$\sin^2 2 \theta_{13}$ versus the smallest mass 
for a normal mass ordering (right). For the first plot we 
have also given the general upper and lower limit 
of the effective mass when the currently allowed $3\sigma$ 
values of the oscillation parameters are used. The corresponding 
plots for category (ib) look basically identical.}
\end{center}
\end{figure}
\begin{figure}[ht]
\begin{center}
\epsfig{file=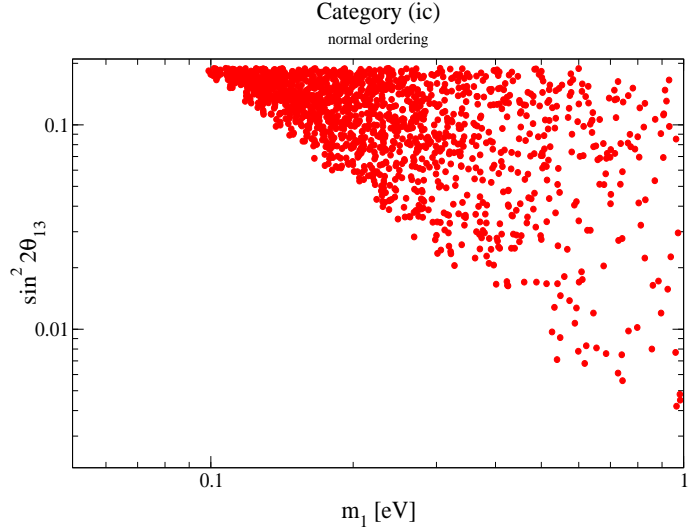,width=9cm,height=7cm}
\caption{\label{fig:1ca}Category (ic) or $(m_\nu)_{23}=0$: 
scatter plot of $\sin^2\theta_{13}$ versus 
the smallest mass in  
case of a normal mass ordering.}
\end{center}
\end{figure}

Let us next discuss the neutrino mixing properties of category (i), in 
which case the low energy neutrino mass matrix contains a vanishing 
off-diagonal entry. 
The phenomenology of mass matrices $m_\nu$ having 
one single texture zero was analyzed in Ref.~\cite{MR}. 
Consider first the case of $(m_\nu)_{12} = 0$, i.e., category (ia). 
The first thing to note is that zero $U_{e3}$ is incompatible 
with the condition of vanishing $(m_\nu)_{12}$ \cite{MR}. 
To elaborate, expanding in terms of $|U_{e3}|$ one finds 
\bea \D \label{eq:m12}
(m_\nu)_{12} = \left(m_2 \, e^{2 i \alpha} - m_1 \right) 
\cos \theta_{12} \, \sin \theta_{12} \, \cos \theta_{23} \\ \D
+ e^{i \delta} \, \sin \theta_{23} \left( 
e^{2 i \beta} \, m_3 - \sin^2 \theta_{12} \, 
m_2 \, e^{2 i \alpha} - \cos^2 \theta_{12} \, m_1 
\right) \, |U_{e3}|
\, ,
\eea
plus higher order terms of $|U_{e3}|$. 
The magnitude of the bracketed part of the zeroth order term is 
bounded from below roughly by 
$\sqrt{\dms}$ for a normal hierarchy, by 
$\frac 12 \dms/\sqrt{\dma}$ for an inverted hierarchy and by 
$\frac 12 \dms/m_0$ for quasi-degenerate neutrinos with an average mass
$m_0$. 
The 12-element therefore cannot vanish for $U_{e3} = 0$, but 
can vanish if $U_{e3}$ departs from zero. It is clear from 
the above expression that in case of an inverted hierarchy and 
for quasi-degenerate neutrinos the bracketed part of the zeroth 
order term in Eq.~(\ref{eq:m12}) has to be small in order to allow the 
term of order $\theta_{13}$ to cancel it. This in turn means that 
$\sin \alpha$ has to be close to zero. In this case there are 
almost no cancellations in the effective mass 
$\meff = |(m_\nu)_{11}|$ governing \onbb$\!\!$ and we have 
\be 
\baz 
\meff \simeq  \sqrt{\dma} \, \cos^2 \theta_{13}  
& \mbox{inverted hierarchy\,,}\\ 
m_0\cos^2 \theta_{13} \le \meff \le m_0 & \mbox{quasi-degeneracy\,.} 
\ea
\ee
The above value and range 
have to be compared with the lower limits, arising from maximal effects of 
Majorana phases, i.e., $\meff_{\rm min} \simeq
\cos^2 \theta_{13} \, \cos 2 \theta_{12}\, \sqrt{\dma} $ 
and $\meff_{\rm min} \simeq \cos^2 \theta_{13} \, \cos 2 \theta_{12}\, m_0$, 
respectively. 
The left panel of Fig.~\ref{fig:1a} shows for category (ia) 
a scatter plot of  the effective mass versus the smallest mass in 
case of an inverted mass ordering. It is clearly seen that 
the largest possible values of \meff~are mostly populated. 
We also show in the right panel a scatter plot of 
$\sin^2 2 \theta_{13}$ versus the smallest mass 
for a normal mass ordering. 
In order to generate this and other plots to be presented later, we 
have varied all neutrino parameters within their allowed 
$3\sigma$-ranges quoted in Eq.~(\ref{eq:data}). 
The results for category (ib), i.e., $(m_\nu)_{13} = 0$, 
are basically identical to the ones for category (ia) \cite{MR}. 
Formally one can obtain the 13-entry of $m_\nu$ from the 12-entry by 
replacing in the latter $\sin \theta_{23}$ with $\cos \theta_{23}$ 
and  $\cos \theta_{23}$ with $-\sin \theta_{23}$.

\begin{figure}[ht]
\begin{center}
\epsfig{file=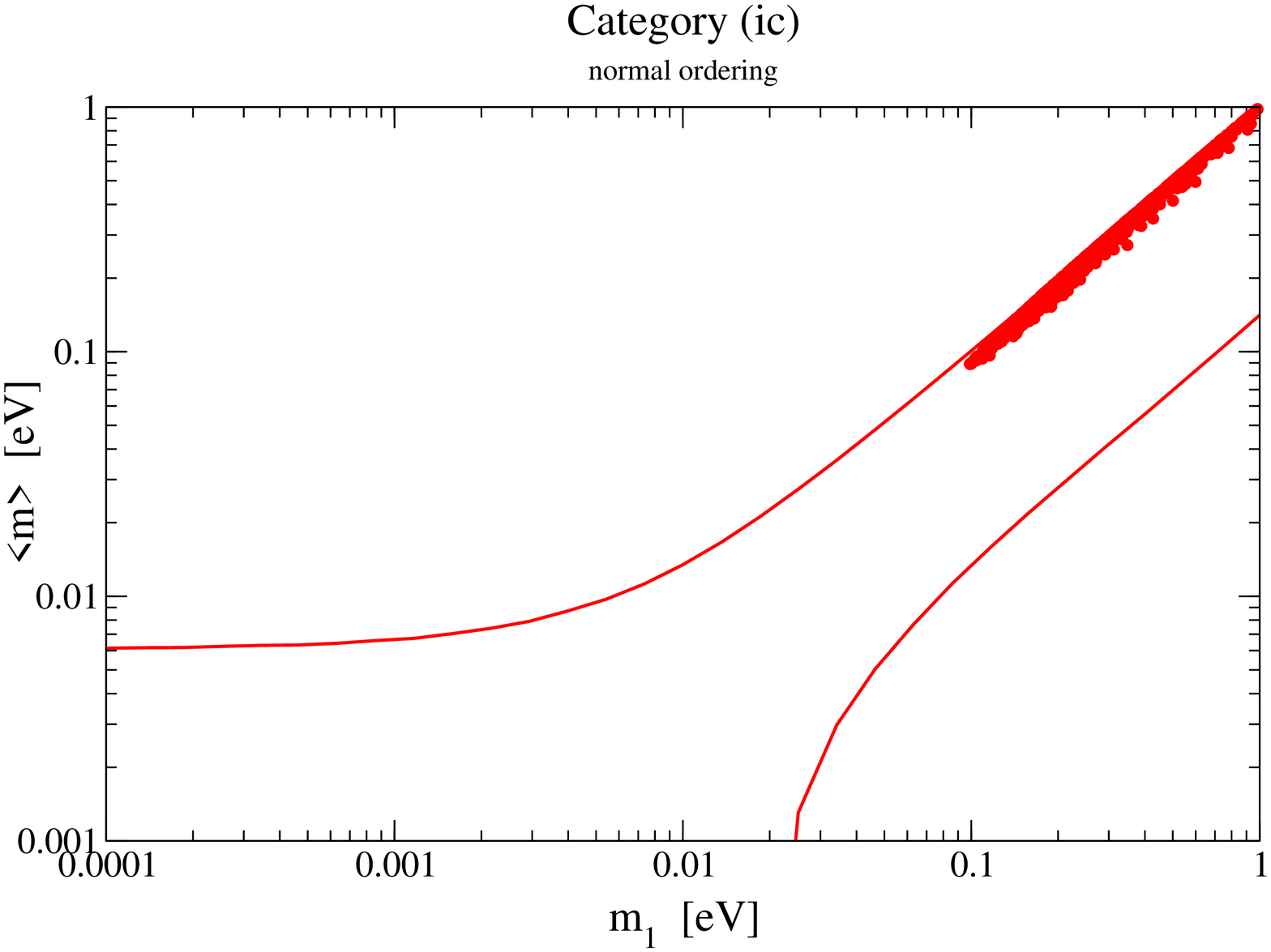,width=8cm,height=8cm}
\epsfig{file=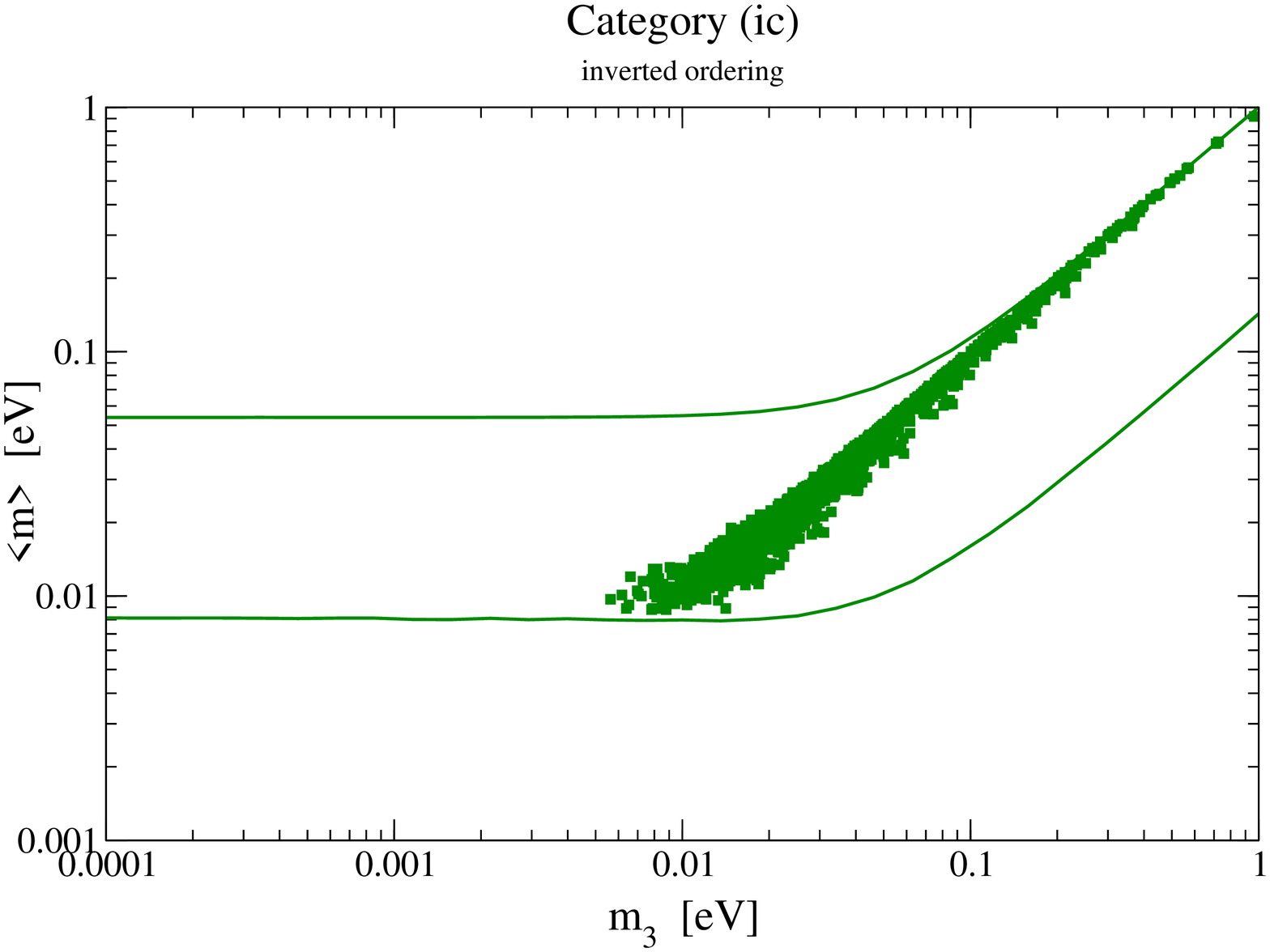,width=8cm,height=8cm}
\caption{\label{fig:1cb}Category (ic) or $(m_\nu)_{23}=0$: 
scatter plots of the 
effective mass versus the smallest 
in case of a 
normal mass ordering (left) and an inverted ordering (right). We 
have also given the general upper and lower limits  
of the effective mass when the currently allowed $3\sigma$ 
values of the oscillation parameters are used. }
\end{center}
\end{figure}

Turning to category (ic), it can be shown that, for 
$\theta_{13} = 0$ and a normal mass ordering, the 23-element 
of $m_\nu$ cannot vanish unless neutrino masses are 
above several eV \cite{MR}. Fig.~\ref{fig:1ca} illustrates this 
by showing the distribution of the smallest mass against 
$\sin^2 2 \theta_{13}$ in case of a normal mass ordering. 
For quasi-degenerate neutrinos, we can express the 
mass matrix element as 
\[ 
(m_\nu)_{23} 
\simeq -m_0 \, \cos \theta_{23} \, \sin \theta_{23} 
\left( 
(e^{2 i \alpha} \, \cos^2 \theta_{12} + \sin^2 \theta_{12}) 
- e^{2i(\beta + \delta)}
\right)\, ,
\] 
where for simplicity we have also set $\theta_{13}$ to zero. 
It is clear that, in order to make $(m_\nu)_{23} $ vanish, 
the expression in the brackets should be very small, or that the relations 
$\sin \alpha \simeq \sin(\beta + \delta) \simeq0$ should hold. This leads 
again to little cancellation in the effective mass, as is obvious from 
Fig.~\ref{fig:1cb}.\\

We close this Section by commenting on the possibility that more than 
one entry of the low energy mass matrix vanishes. In this respect it is known 
that, in the mass diagonal charged lepton basis, two is the maximum 
number of vanishing elements allowed 
in the neutrino mass matrix \cite{2zero}. 
We refer to Ref.~\cite{2zero} for details on 
the phenomenology of these cases. 
It is sufficient to note here that seven of those cases exist, 
namely the simultaneous vanishing of the 11- and 12-, the 11- and 13-, 
the 12- and 22-, the 13- and 22-, the 13- and 33-, the 12- and 33-, 
and finally the 22- and 33-entries. 
In category (ia), in which $(m_\nu)_{12} = 0$, there is the possibility 
that in addition $(m_\nu)_{11}$, $(m_\nu)_{22}$ or $(m_\nu)_{33}$ can be zero. 
This in turn means that the branching ratios of the decays 
$\tau \ra e \gamma$ and $\tau \ra \mu \gamma$, which depend on 
$|(m_\nu)_{13}|^2$ and $|(m_\nu)_{23}|^2$, respectively, are guaranteed 
to be non-zero. The same is true for category (ib), in which case 
$(m_\nu)_{13} = 0$, and for which again only $(m_\nu)_{11}$, $(m_\nu)_{22}$ or 
$(m_\nu)_{33}$ can be zero. BR($\mu \ra e \gamma$) and 
BR($\tau \ra \mu \gamma$) then are non-zero because there are 
proportional to the non-zero $|(m_\nu)_{12}|^2$ and $|(m_\nu)_{23}|^2$, 
respectively. In case of category (ic), or $(m_\nu)_{23} = 0$, it 
turns out that no other mass matrix element can vanish, and therefore 
BR($\mu \ra e \gamma$) and BR($\tau \ra e \gamma$) are necessarily 
non-zero.

\subsection{\label{sec:2}Category (ii)}
Now, let us similarly consider the subclass of category (ii) in 
which columns $l$ and $k~(\neq l)$ are orthogonal, 
element by element. 
For this subclass of four zero textures, the relation 
\be
(m_D^\dagger \, m_D)_{lk} = 0~
\label{eq:ybcat2}
\ee
applies. This means that the terms ${\cal I}_{lk}^\alpha$ and 
$\varepsilon_l$ vanish in the respective flavored and 
unflavored heavy neutrino decay asymmetries, 
cf.~Eqs.~(\ref{eq:epsIal}) and (\ref{eq:epsI}). \\

It will be illuminating to explicitly see the different seesaw 
induced physical  
effects that appear in a given subclass of textures. Table 
\ref{tab:2a} has been made for this purpose by considering 
category (iia). 
The leftmost column contains the Dirac mass matrices for particular textures 
of this category after one phase per row has been rotated away. 
This is followed in subsequent columns by 
expressions for only those heavy neutrino decay asymmetries which are 
nonzero and contribute accordingly to leptogenesis 
as well as their corresponding wash-out factors. 
For simplicity, we work as before under the 
assumption that $M_1 \ll M_{2,3}$ so that one needs to 
consider only the decay of $N_1$, i.e., $i$ is always 1. 
However, this is an inessential assumption. 
Our tables can easily be generalized to the
cases with
$j \neq i$.
Interestingly, we find again that 
the ${\cal J}_{1j}^\alpha$ are all zero.
The entries in the last 
column of Table 
\ref{tab:2a} are expressions describing the invariant $J_{\rm CP}$ 
relevant to CP violation that can be observed in neutrino oscillations. 
The latter is always the sum of three terms and we have chosen 
not to list common proportionality factors. 
Again, in analogy to the discussion in the previous Subsection, 
a comparison of the entries in the leptogenesis and the
$J_{\rm CP}$ columns is instructive. It shows how there could be 
a one-to-one correspondence between the leptogenesis phase and 
the low energy Dirac phase $\delta$. For instance, consider 
the first row in Table \ref{tab:2a}, for which 
only $\varepsilon_1^\tau$ contributes 
to leptogenesis, and for which the CP phase denoted by $\gamma_1$ is crucial. 
The same phase 
can control low energy (Dirac) CP violation 
provided the condition 
\be
b_2^2 \, M_1 \, M_3 \gg |c_1^2 \, M_2 \, M_3 \, \sin 2 \beta_2|\, , 
|(a_3^2 + b_3^2 + c_3^2) \, M_1 \, M_2 \, \sin 2(\beta_2 - \gamma_1)|
\ee
is fulfilled. 
In the last two rows of Table \ref{tab:2a} there are two nonzero 
decay asymmetries, which lead to more possibilities. 
In this respect, one may note that hierarchical heavy neutrinos 
lead to a suppression of ${\cal I}_{13}^\alpha$ with respect to 
${\cal I}_{12}^\alpha$ by a factor $M_2/M_3$. 
Tables \ref{tab:2b} and \ref{tab:2c} show the leptogenesis related 
phenomenology of categories (iib) and (iic), respectively. 
The matrices can be obtained from the ones of category (iia) 
by interchanging the first with the second 
(first with the third) row. \\ 

Even though none 
of the branching ratios is guaranteed to 
vanish in this class of textures, there 
exists an interesting feature. For all 18 matrices in 
category (ii) the following relation applies for $i \neq j$: 
\be 
|(m_D \, L \, m_D^\dagger)_{ij}| \propto |(m_\nu)_{ij}|\, ,
\ee
i.e., the branching ratios 
for the decays $\ell_i \ra \ell_j \gamma$ are proportional 
to the square of the 
modulus of the $ij$ element of the low energy mass matrix. 
This is similar to the situation in category (i). The crucial 
difference is however that in category (ii) the ratios of 
branching ratios are related to ratios of neutrino mass matrix elements 
without any ambiguity coming from the unknown values of the heavy Majorana 
neutrino masses. Consider again the example from 
the first row of Table \ref{tab:2a}. The low energy Majorana mass matrix is 
given as 
\be \D
m_\nu = -
\left(
\bad
\D \frac{a_3^2}{M_3} \D & \D \frac{a_3 b_3 }{M_3}  \D  & \D   
\frac{a_3 c_3}{M_3} \\
\cdot &  \D \frac{b_2^2 e^{2 i \beta_2}}{M_2} + \frac{b_3^2}{M_3}  & 
 \D \frac{b_3 c_3 }{M_3} \\
\cdot & \cdot &  \D \frac{c_1^2 e^{2i \gamma_1}}{M_1} 
+ \frac{c_3^2}{M_3} 
\ea
\right)\,.
\ee
The relevant expressions for LFV are found to be  
\[ 
\left|(m_D \, L \, m_D^\dagger)_{12}\right|^2 
= a_3^2 \, b_3^2 \, L_3^2 ~,~~
\left|(m_D \, L \, m_D^\dagger)_{13}\right|^2 
= a_3^2 \, c_3^2 \, L_3^2
~~\mbox{ and }~
\left|(m_D \, L \, m_D^\dagger)_{23}\right|^2 
= b_3^2 \, c_3^2 \, L_3^2\,.
\] 
Therefore, the ratios of branching ratios are unambiguously 
given by the ratios of the corresponding mass matrix elements. 
\begin{figure}[ht]
\begin{center}
\epsfig{file=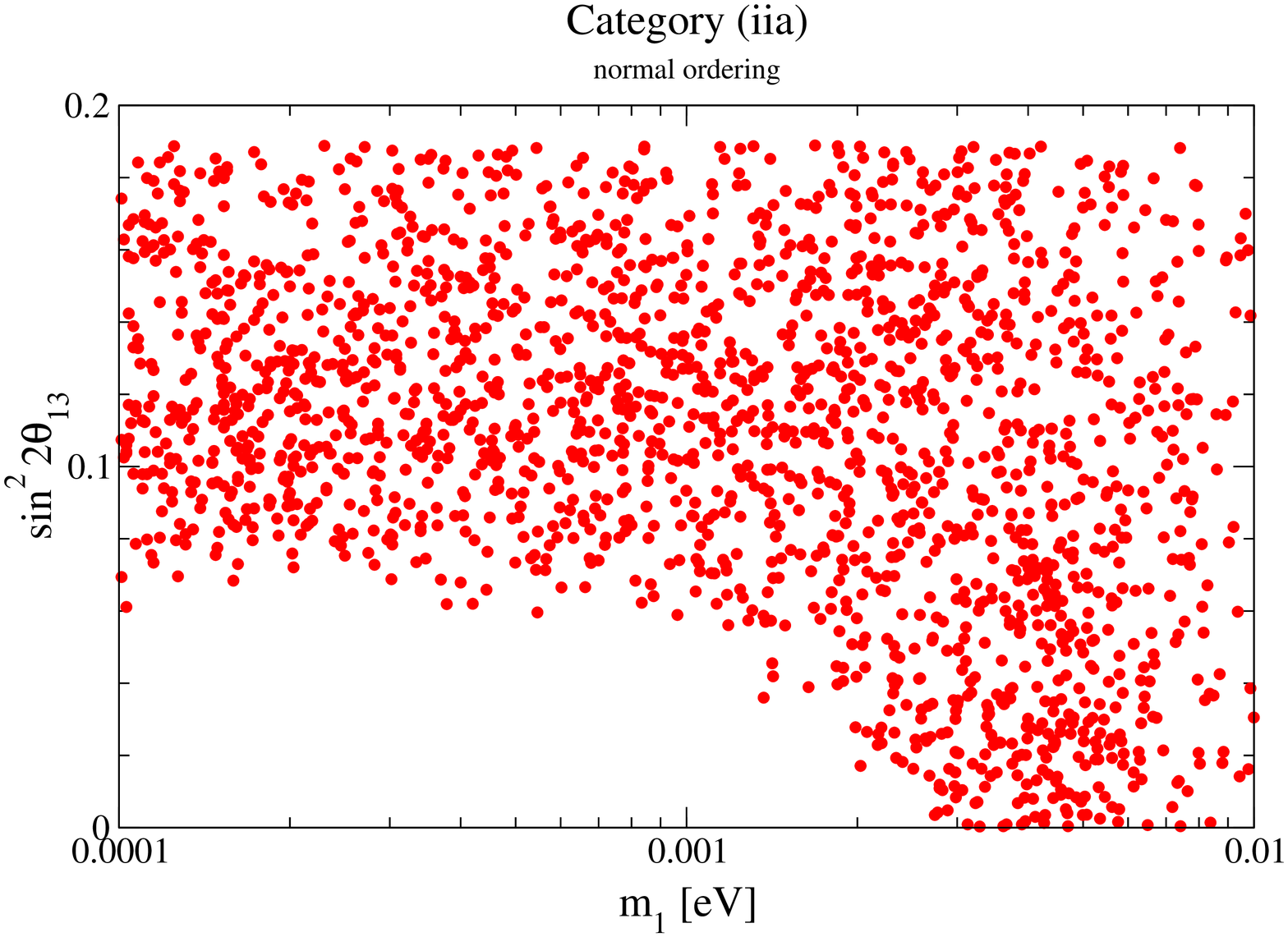,width=8cm,height=8cm}
\epsfig{file=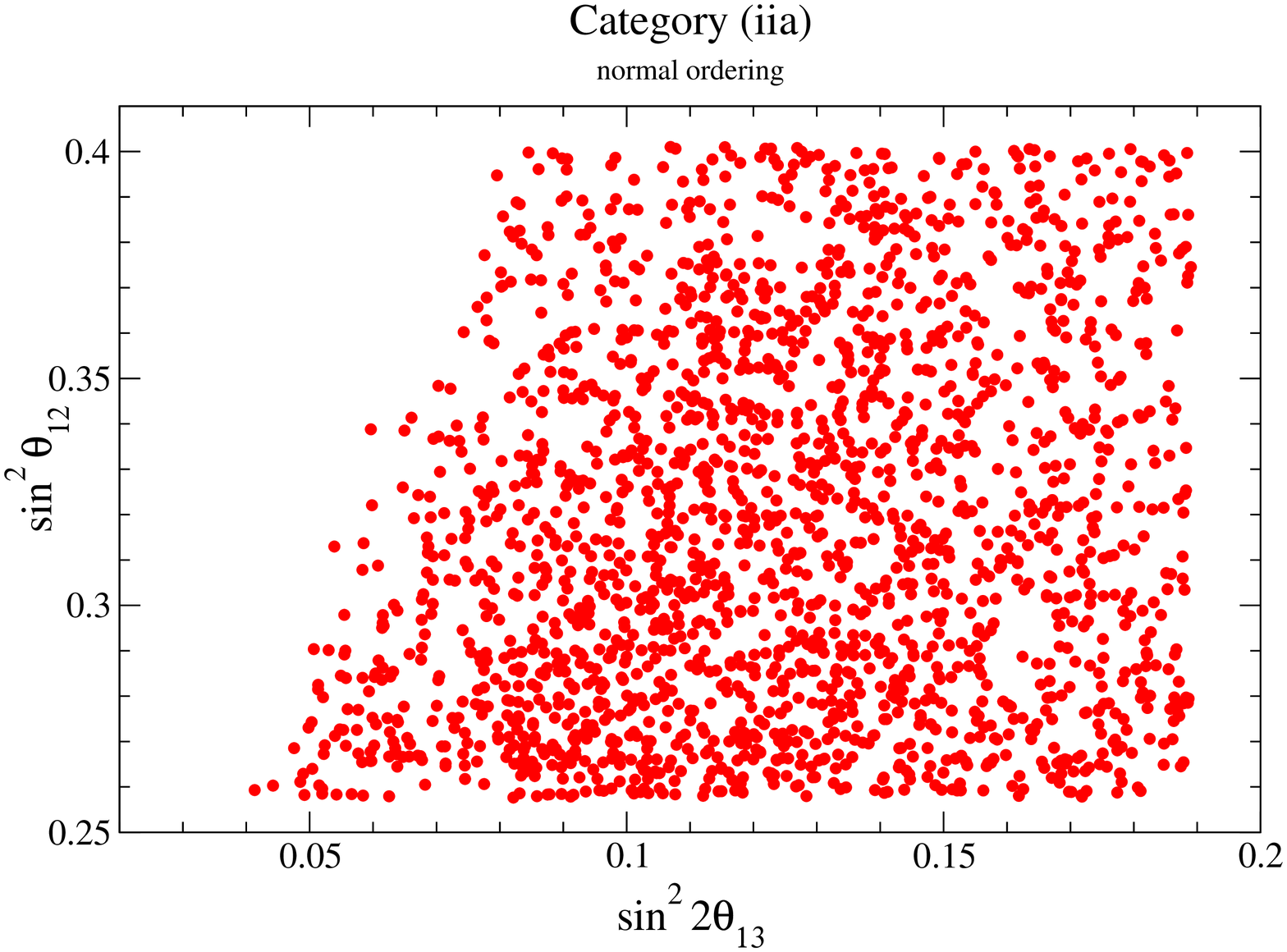,width=8cm,height=8cm}
\caption{\label{fig:2a}Category (iia): 
scatter plots of $\sin^2 2 \theta_{13}$ versus the smallest mass $m_1$ 
and of $\sin^2 2 \theta_{12}$ versus 
$\sin^2 \theta_{13}$ (for $m_1 = 0.001$ eV) in 
case of a normal mass ordering and when the conditions  
$|(m_\nu)_{11} \, (m_\nu)_{23}| = |(m_\nu)_{12} \, (m_\nu)_{13}|$ and 
Im$\{(m_\nu)_{11} \, (m_\nu)_{23} 
\, (m_\nu)_{12}^\ast \, (m_\nu)_{13}^\ast\} = 0$ 
are fulfilled.}
\end{center}
\end{figure}
This ``minimal lepton flavor violation'' scenario 
in principle allows one to 
predict the rates from  measurable low energy mass matrix elements. 
Indeed, from Eq.~(\ref{eq:branching}), we see that  
the following relations now hold in category (iia): 
\bea \D \label{eq:BRvs.mnu}
\frac{{\rm BR}(\tau \ra e \gamma)}{{\rm BR}(\tau \ra \mu \gamma)} 
\simeq \left|\frac{(m_\nu)_{13}}{(m_\nu)_{23}}\right|^2
\, ,\\[0.2cm] \D 
\frac{{\rm BR}(\mu \ra e \gamma)}{{\rm BR}(\tau \ra e \gamma)} 
\simeq \frac{1}{{\rm BR}(\tau \ra e \, \nu \overline{\nu})} 
\left|\frac{(m_\nu)_{12}}{(m_\nu)_{13}}\right|^2 
\,. 
\eea 
Eq.~(\ref{eq:BRvs.mnu}) is valid for 
all textures of category (ii).\\

\begin{figure}[ht]
\begin{center}
\epsfig{file=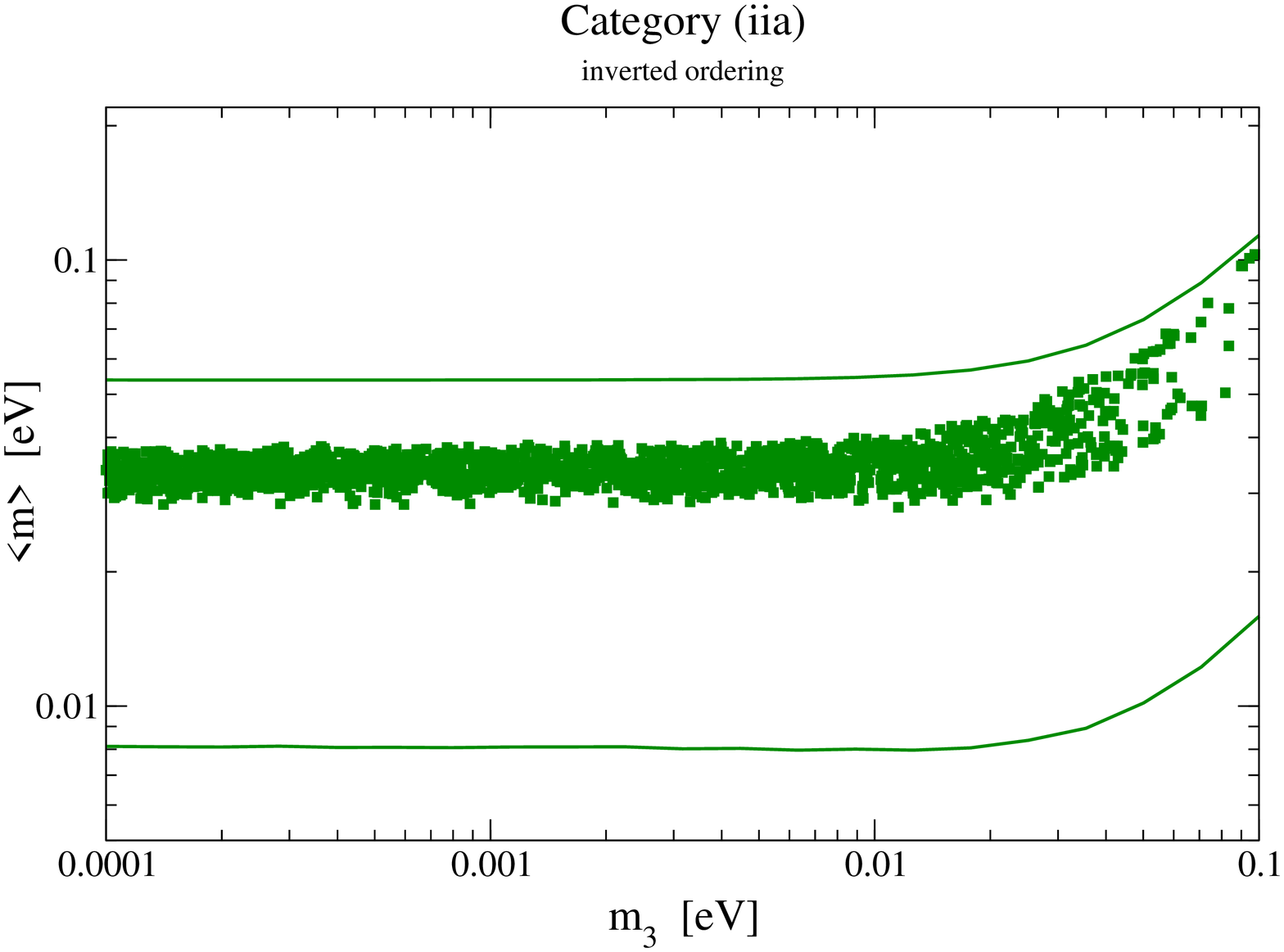,width=9cm,height=7cm}
\caption{\label{fig:2aIH}Category (iia): 
scatter plot of the effective mass $\meff$ versus the smallest mass $m_3$ in 
case of an inverted  mass ordering and when the conditions  
$|(m_\nu)_{11} \, (m_\nu)_{23}| = |(m_\nu)_{12} \, (m_\nu)_{13}|$ and 
Im$\{(m_\nu)_{11} \, (m_\nu)_{23} 
\, (m_\nu)_{12}^\ast \, (m_\nu)_{13}^\ast\} = 0$ 
are fulfilled. The solid (green) lines correspond to the upper and 
lower limit of the effective mass when the currently allowed 
$3\sigma$ ranges of the oscillation parameters are used.}
\end{center}
\end{figure}

The three sub-categories in category (ii) 
have in addition correlations between the low energy mass 
matrix elements, given in Eqs.~(\ref{eq:2a}, \ref{eq:2b}, \ref{eq:2c}), 
which lead to 
\be \label{eq:ratio1}
\frac{{\rm BR}(\tau \ra e \gamma)}{{\rm BR}(\tau \ra \mu \gamma)} 
\simeq \left|\frac{(m_\nu)_{13}}{(m_\nu)_{23}}\right|^2 
= \left\{ \baz 
\D \left| \frac{(m_\nu)_{11}}{(m_\nu)_{12}}\right|^2
 & \mbox{category (iia)}\, , \\[0.2cm]
\D \left| \frac{(m_\nu)_{12}}{(m_\nu)_{22}}\right|^2  
 & \mbox{category (iib)}\, , \\[0.2cm]
\D \left| \frac{(m_\nu)_{12} \, (m_\nu)_{33}}{(m_\nu)_{23}^2}\right|^2 
& \mbox{category (iic)}~
\ea \right. 
\ee  
and 
\be\label{eq:ratio2}
\frac{{\rm BR}(\mu \ra e \gamma)}{{\rm BR}(\tau \ra e \gamma)} 
\, {\rm BR}(\tau \ra e \, \nu \overline{\nu}) 
\simeq \left|\frac{(m_\nu)_{12}}{(m_\nu)_{13}}\right|^2 
= \left\{ \baz 
\D \left| \frac{(m_\nu)_{11} \, (m_\nu)_{23}}{(m_\nu)_{13}^2}\right|^2
 & \mbox{category (iia)}\, , \\[0.2cm]
\D \left| \frac{(m_\nu)_{22}}{(m_\nu)_{23}}\right|^2  
 & \mbox{category (iib)}\, , \\[0.2cm]
\D \left| \frac{(m_\nu)_{23}}{(m_\nu)_{33}}\right|^2 
& \mbox{category (iic)}\,.
\ea \right. 
\ee

Because the neutrino mass matrix obeys $\mu$--$\tau$ (or 2--3) 
symmetry to a good approximation, $|(m_\nu)_{12}/(m_\nu)_{13}|^2$  
is typically 1. Therefore, BR$(\tau \ra e \gamma) 
\simeq 0.178 \, {\rm BR}(\mu \ra e \gamma)$. 
With the current upper limit of $1.2 \times 10^{-11}$ 
on BR$(\mu \ra e \gamma)$, and an expected improvement of 
at most two 
orders of magnitude on the 
limit of ${\rm BR}(\tau \ra e \gamma) \le 1.1 
\times 10^{-7}$, it follows that 
in this scenario $\tau \ra e 
\gamma$ will probably not be observed in a foreseeable future. 
Since it turns out that BR$(\tau \ra \mu \gamma) 
\sim {\rm BR}(\tau \ra e \gamma)$, the same is true 
for the decay $\tau \ra \mu \gamma$.
To discuss further the phenomenology of category (iia), 
we show in Fig.~\ref{fig:2a} scatter plots for a 
normal mass ordering in case 
the correlation (\ref{eq:2a}) holds. 
Obtaining analytical correlations for this example 
is very complicated. 
If we set $m_1 = \theta_{13} = 0$ and 
$\theta_{23} = \pi/4$, then Eq.~(\ref{eq:2a}) leads to 
$\frac 12 \, 
\sqrt{\dma/\dms} = \cos^2 \theta_{12} \, \cos 2(\alpha - \beta - \delta) 
< 1$, which is incompatible with the available data. This implies 
the presence of a lower limit on 
$\sin^2 2 \theta_{13}$, which indeed 
turns out to be roughly 0.04~(0.06) for 
$m_1 =0.001~(0)$ eV. 
This lower limit vanishes when the neutrino mass increases, 
which can clearly be seen in Fig.~\ref{fig:2a}.\\

With regard to an inverted mass hierarchy, we find no correlation 
for the neutrino oscillation parameters. Instead, 
the effective mass $\meff = |(m_\nu)_{11}|$ governing 
neutrinoless double beta decay is constrained by 
the correlation in Eq.~(\ref{eq:2a}). 
Note that in the definition of the correlation for 
category (iia) the effective mass ($(m_\nu)_{11}$) appears 
explicitly. The influence on \meff~can be estimated 
by considering the equality 
$|(m_\nu)_{11} \, (m_\nu)_{23}|^2 = |(m_\nu)_{21} \, (m_\nu)_{13}|^2$ 
in an approximate manner. Neglecting 
$m_3$, setting $m_2 \simeq m_1 \simeq \sqrt{\dma}$, 
and inserting $\sin^2 \theta_{23} = \frac 12$, we obtain from it 
the condition 
\[ 
\frac{(\dma)^2}{4} \left(1 - 2 \sin^2 2 \theta_{12} \, 
\sin^2 \alpha \right) + {\cal O}(|U_{e3}|^2) 
\stackrel{!}{=} 0 \,.
\] 
Hence, $\sin^2 \alpha \simeq 1/(8\, \sin^2 \theta_{12} \,\cos^2 \theta_{12})$ 
(which is equal to $\frac{9}{16}$ if $\sin^2 \theta_{12} = \frac 13$) 
and inserting this in the effective mass leads to
\be \label{eq:meffiia}
\meff \simeq 
\frac{\cos^2 \theta_{13}}{\sqrt{2}} \sqrt{\dma}\,. 
\ee
This has to be compared with the general lower and 
upper limits on \meff, which are 
$\cos 2 \theta_{12} \, \cos^2 \theta_{13} \,\sqrt{\dma}$ 
and $\cos^2 \theta_{13} \,\sqrt{\dma}$, respectively. 
Fig.~\ref{fig:2aIH} 
shows how the effective mass as a function of the smallest mass 
$m_3$ has considerably less spread than without the 
correlation Eq.~(\ref{eq:2a}). 
For the other two conditions in Eqs.~(\ref{eq:2b}) and 
(\ref{eq:2c}) we did not find any interesting correlations between 
the neutrino observables.\\

\begin{figure}[tb]
\begin{center}
\epsfig{file=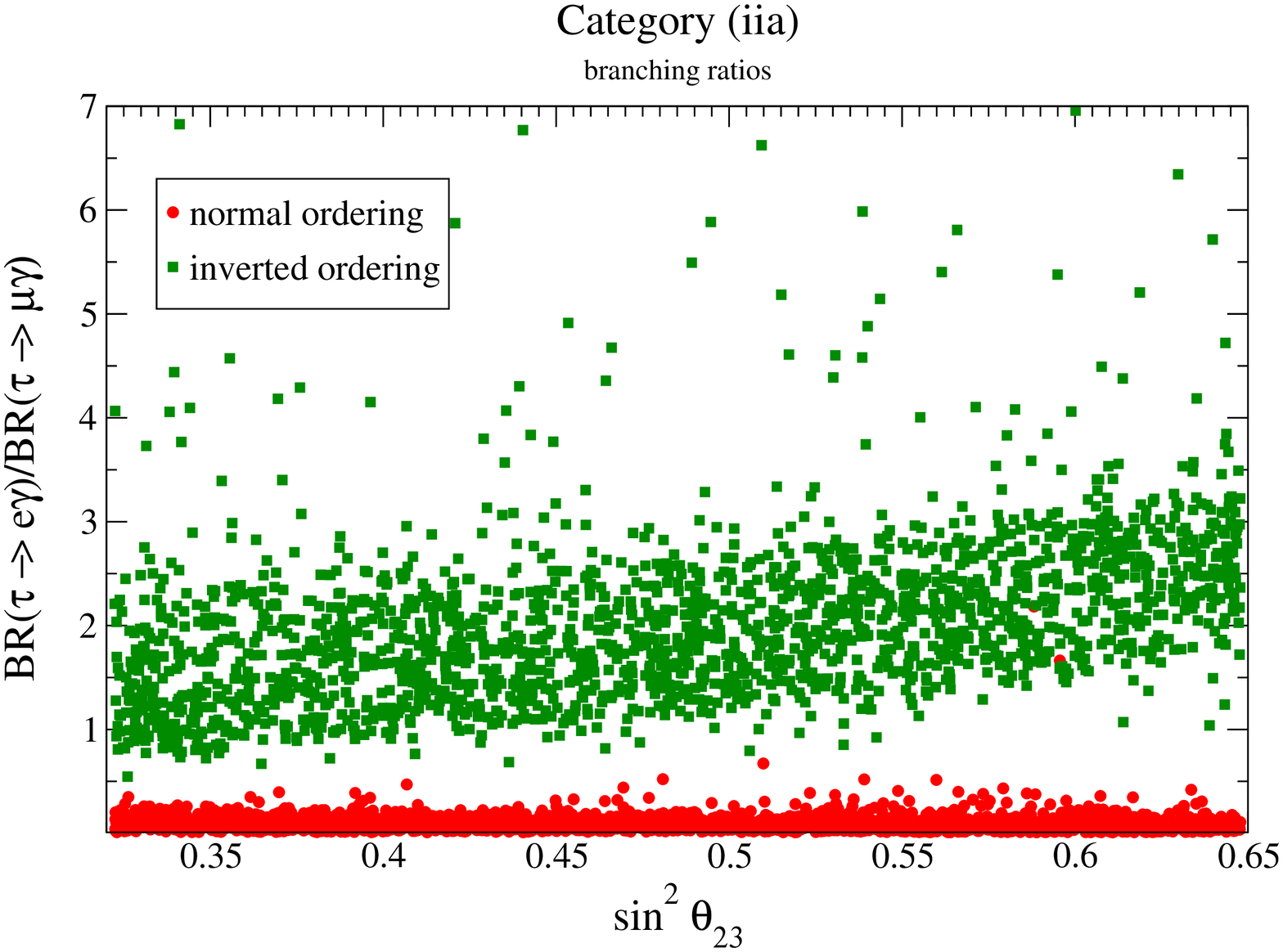,width=8cm,height=8cm}
\epsfig{file=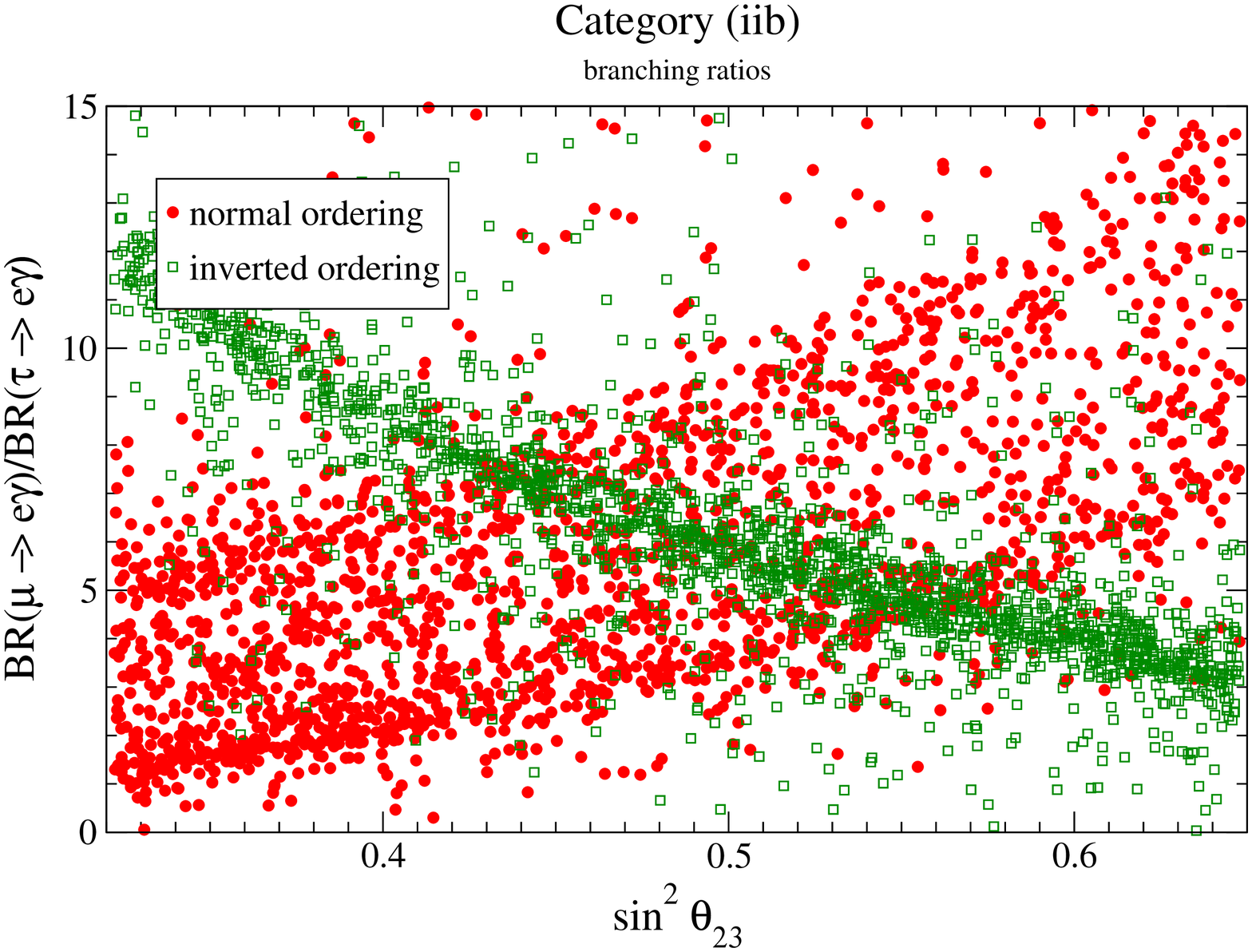,width=8cm,height=8cm}
\caption{\label{fig:BR2a}Category (iia): 
scatter plots of BR$(\tau \ra e \gamma)$/BR$(\tau \ra \mu \gamma)$ 
(left) and  BR$(\mu \ra e \gamma)$/
BR$(\tau \ra e \gamma)$ (right) against 
$\sin^2 2 \theta_{23}$ when the conditions  
$|(m_\nu)_{11} \, (m_\nu)_{23}| = |(m_\nu)_{12} \, (m_\nu)_{13}|$ and 
Im$\{(m_\nu)_{11} \, (m_\nu)_{23} 
\, (m_\nu)_{12}^\ast \, (m_\nu)_{13}^\ast\} = 0$ 
are fulfilled. The red circles are for the normal ordering, the 
green squares for the inverted one. }
\end{center}
\end{figure}

Turning to the ratios of branching ratios, Eqs.~(\ref{eq:ratio1}) 
and (\ref{eq:ratio2}) are phenomenologically very interesting. 
In Fig.~\ref{fig:BR2a} we show as a function of $\sin^2 \theta_{23}$,
the ratio of the 
branching ratios of $\tau \ra e \gamma$ 
and $\tau \ra \mu \gamma$ in the left panel 
and of $\mu \ra e \gamma$ and $\tau \ra e \gamma$ 
in the right panel.
These numbers are equal to 
$|(m_\nu)_{13}|^2/|(m_\nu)_{23}|^2$ and 
$(|(m_\nu)_{12}|^2/|(m_\nu)_{13}|^2) /{\rm BR}(\tau 
\ra e \, \overline{\nu}  \nu)$, respectively. Note that, for  
normal ordering, the $\mu$--$\tau$ block of $m_\nu$ is usually 
larger than the elements of the $e$-row, while for the 
inverted ordering case, all elements of $m_\nu$ are of similar 
magnitude. This explains why in the left panel the predictions 
for the inverted ordering are higher compared to that for the 
normal ordering. In addition, since 
${\rm BR}(\tau \ra e \, \overline{\nu}  \nu) = 0.178$, 
the ratios shown in the right panel are larger 
than those in the left panel. 
To be more specific, let us consider 
a simplified example. For inverted ordering, assuming 
$\theta_{13} = m_3 = 0$, $m_2 \simeq m_1$ and $\sin^2 
\theta_{12} = \frac 13$, we obtain
\[ 
\frac{|(m_\nu)_{13}|^2}{|(m_\nu)_{23}|^2} \simeq 
\frac{1}{\cos^2 \theta_{23}}\, ,
\]
which explains the mild increase of this ratio as a function of 
$\sin^2 \theta_{23}$ in the left panel of Fig.~\ref{fig:BR2a}. 
On the other hand, 
the ratio BR$(\mu \ra e \gamma)$/BR$(\tau \ra e \gamma)$ is seen to 
be proportional to $\cot^2 \theta_{23}$ for the same set of 
assumptions. This therefore results in the decrease 
seen for inverted ordering in the right panel of Fig.~\ref{fig:BR2a}. 
For normal ordering and with the same set of assumptions 
one can show that 
$|(m_\nu)_{12}|^2/|(m_\nu)_{13}|^2 \propto \tan^2 \theta_{23}$. 
This explains the increase of 
BR$(\mu \ra e \gamma)$/BR$(\tau \ra e \gamma)$ with 
$\sin^2 \theta_{23}$ in the right panel of 
Fig.~\ref{fig:BR2a}. 
Note that, for the inverted mass ordering 
BR$(\tau \ra e \gamma)$$\sim$BR$(\tau \ra \mu \gamma)$, while 
BR$(\mu \ra e \gamma)$$\sim$BR$(\tau \ra e \gamma)$ is 
true irrespective of the neutrino mass spectrum. 
Therefore, we do not expect BR$(\tau \ra e \gamma)$ to be 
observed in the forthcoming experiments \cite{BR_fut}. 
Similarly for the inverted ordering, 
the predicted BR$(\tau \ra \mu \gamma)$
is not expected to be checked experimentally in the 
next generation experiments \cite{BR_fut}. 
The scatter plots of 
BR$(\mu \ra e \gamma)$/BR$(\tau \ra e \gamma)$ in all sub-categories 
of category (ii) look very similar and the other cases show no 
interesting correlations.\\

Finally, we discuss the possibility and implications of vanishing 
branching ratios, and hence the 
low energy mass matrix elements. Since the basic 
criteria for category (ii) comprise of zero sub-determinant 
conditions (cf. Eq. (\ref{eq:2a})-(\ref{eq:2c})), one can
easily note that in this case, the vanishing of one 
element of $m_\nu$ would necessitate the vanishing of 
another mass matrix element as well. Therefore, for category (ii) 
we cannot have just one zero $m_\nu$ texture, and hence just one vanishing 
branching ratio. The allowed two zero textures 
of $m_\nu$ have been extensively studied in the 
literature \cite{2zero}. One can check that none of 
the phenomenologically viable two zero textures allow 
$(m_\nu)_{23}$ to be zero. An immediate consequence of this is 
that for category (ii), BR$(\tau \ra \mu \gamma)\neq 0$. 
Among the sub-categories, we note that for category (iia), 
if BR$(\mu \ra e \gamma)=0$ (implying $(m_\nu)_{12}=0$), 
then $(m_\nu)_{11}=0$. Therefore, for this case a 
vanishing BR$(\mu \ra e \gamma)$ predicts vanishing 
neutrinoless double beta decay -- this in turn is possibile  
only for normal mass ordering with $m_1 \sim 0.005$ eV. 
For category (iib), the only allowed two zero texture is 
with $(m_\nu)_{12}=0$ and $(m_\nu)_{22}=0$. For this category 
a vanishing  $(m_\nu)_{13}$ would make $(m_\nu)_{12}=0$ and this 
case is strongly disfavored by the data. Therefore, 
BR$(\tau \ra e \gamma)\neq 0$ for this case while 
BR$(\mu \ra e \gamma)$ could go to zero. 
On the other hand, for category (iic), the $(m_\nu)_{12}\neq 0$ 
condition is imposed by the data, implying that 
BR$(\mu \ra e \gamma)\neq 0$, while 
BR$(\tau \ra e \gamma)$ could go to zero if both 
$(m_\nu)_{13}=0$ and $(m_\nu)_{22}=0$ simultaneously.

\section{\label{sec:concl}Summary and Conclusions }

We have looked at phenomenological constraints from the two maximally allowed 
categories of four zero textures in $Y_\nu$ in the basis in 
which the mass matrices 
$M_R$ and $m_\ell$ are diagonal. Our framework is that 
of the supersymmetric type I 
seesaw and we have examined the consequences on seesaw parameters 
of the conditions 
imposed on elements of the neutrino Majorana mass matrix $m_\nu$ by 
either category 
of textures. We have included the effective Majorana mass 
$\langle m \rangle$, appearing in 
neutrinoless nuclear double beta decay, in the list of 
seesaw parameters studied. 
Our use of those conditions is reliable in that the latter 
are radiatively stable, being invariant under RG running. \\

We have demonstrated via various scatter plots how restricted 
regions in the seesaw parameter space are selected by the said conditions. 
For the textures of each category, we have further derived a 
number of results on radiative LFV decays, several of them with 
observable consequences. For instance, any observation of the 
decay $\ell_i \ra \ell_j \gamma$ would rule out those category (i) 
textures which imply that $(m_\nu)_{ij}$ = $(m_\nu)_{ji}$ vanishes. 
In general, all 72 four zero textures predict that the branching ratios 
of  $\ell_i \ra \ell_j \gamma$ are proportional to the absolute values 
squared of the mass matrix element $(m_\nu)_{ij}$ times a function of 
heavy Majorana neutrino masses. For category (ii) 
this function is the same for all $i,j$ 
and the ratios of branching ratios 
are directly given by ratios of low energy mass matrix elements.  
For category (ii), we have been able to generate sample scatter 
plots for ratios of branching ratios such as 
${\rm BR}(\mu \ra e \gamma)$/${\rm BR}(\tau \ra e \gamma)$ and 
${\rm BR}(\tau \ra e \gamma)$/${\rm BR}(\tau \ra \mu \gamma)$ against 
${\rm sin}^2 \theta_{23}$. We have also obtained results with 
physical consequences for leptogenesis from these textures. 
An example is that, for both categories (i) and (ii), we have 
derived conditions which fix the functional form of the leptogenesis 
phase purely in terms of the low energy Dirac phase without 
invoking any other phase from among the seesaw parameters. 
This is always possible for the 
four zero textures because their number of physical phases is two.
We have also provided additional information by 
tabulating quantities relevant to lepton flavor asymmetries, 
wash-out factors and the Jarlskog invariant for each 
texture of category (ii). \\

In conclusion, we have highlighted the rich phenomenological 
structure of the allowed four zero textures in $Y_\nu$ defined in 
the charged lepton and right-handed 
neutrino mass diagonal basis. The consequent conditions on 
$m_\nu$ are radiatively stable. These have been shown to 
lead to significant reductions in the seesaw parameter space. \\

\vspace{0.3cm}
\begin{center}
{\bf Acknowledgments}
\end{center}

We thank Michael Schmidt for helpful discussions. 
S.C.~was supported
by the XI Plan Neutrino Project 
at the Harish-Chandra Research Institute.
W.R.~was supported by the Deutsche Forschungsgemeinschaft 
in the Sonderforschungsbereich 
Transregio 27 ``Neutrinos and beyond -- Weakly 
interacting particles in 
Physics, Astrophysics and Cosmology'' and 
by the ERC under the Starting Grant
MANITOP. P.R.~was supported in 
part by the DAE BRNS of the Government of India and 
acknowledges the hospitality of 
the Harish-Chandra Research Institute. We also thank the 
organizers of WHEPP-X, where this work was initiated, for hospitality.

\pagestyle{empty}

\begin{sidewaystable}
\begin{tabular}{|c|c|c|c|} \hline
$m_D = $ & leptogenesis & wash-out & $J_{CP} \propto $\\ \hline \hline 
$
\left( 
\bad 
0 & 0 & a_3 \\
0 & b_2 e^{i \beta_2} & b_3 \\
c_1 e^{i \gamma_1} & 0 & c_3 
\ea 
\right)$ 
& $\ba {\cal I}_{13}^\tau  = - c_1^2 c_3^2 \sin 2 \gamma_1  \ea 
$
&  $\tilde{m}_1^\tau = c_1^2 /M_1$ 
& 
$ \ba 
b_2^2 {M_1} {M_3} \sin 2 \gamma_1  
- c_1^2 {M_2} {M_3} \sin 2\beta_2 \\ - 
\left(a_3^2 + b_3^2 + c_3^2 \right) {M_1} {M_2} 
\sin 2 ({\beta_2}-{\gamma_1})
\ea $
 \\ \hline 
$
\left( 
\bad 
0 & 0 & a_3 \\
b_1 e^{i \beta_1} & 0 & b_3 \\
0 & c_2 e^{i \gamma_2} & c_3 
\ea 
\right)$ 
& $\ba {\cal I}_{13}^\mu  = -b_1^2 b_3^2 \sin 2 \beta_1 
\ea $  
& $\tilde{m}_1^\mu = b_1^2 /M_1$ 
& 
$\ba 
 b_1^2 {M_2} {M_3} \sin 2 {\gamma_2} - c_2^2 {M_1}
    {M_3} \sin 2 {\beta_1} \\
- \left(a_3^2+b_3^2+c_3^2\right) {M_1}
    {M_2} \sin 2 ({\beta_1}-{\gamma_2}) 
\ea $
\\ \hline
$
\left( 
\bad 
0 & a_2 & 0 \\
0 & b_2 e^{i \beta_2} & b_3 \\
c_1 e^{i \gamma_1} & c_2 & 0 
\ea 
\right)$ 
& $ \ba {\cal I}_{12}^\tau  = -c_1^2 c_2^2 \sin 2 \gamma_1\ea $ 
& $\tilde{m}_1^\tau = c_1^2 /M_1$  
& 
$ \ba 
c_1^2 {M_2} {M_3} \sin 2 {\beta_2} + b_3^2 {M_1}
    {M_2} \sin 2 {\gamma_1} \\ + \left(a_2^2 + 
b_2^2 + c_2^2 \right) {M_1} {M_3} 
\sin 2 ({\beta_2}+{\gamma_1})
\ea $
\\ \hline
$
\left( 
\bad 
0 & a_2 & 0 \\
b_1 & b_2 e^{i \beta_2} & 0 \\
0 & c_2 & c_3 e^{i \gamma_3}  
\ea 
\right)$ 
& ${\cal I}_{12}^\mu  = b_1^2 b_2^2 \sin 2 \beta_2$ 
& $\tilde{m}_1^\mu = b_1^2 /M_1$ 
& 
$\ba 
 b_1^2 {M_2} {M_3} \sin 2 {\gamma_3} + c_3^2 {M_1} 
    {M_2} \sin 2 {\beta_2} \\ + \left(a_2^2 + b_2^2 
+ c_2^2 \right) {M_1} {M_3} \sin 2 ({\beta_2}+{\gamma_3})
\ea $
\\ \hline
$
\left( 
\bad 
a_1 & 0 & 0 \\
b_1 & b_2 e^{i \beta_2} & 0 \\
c_1 & 0 & c_3 e^{i \gamma_3}  
\ea 
\right)$ 
& 
$ \ba 
{\cal I}_{12}^\mu  = b_1^2 b_2^2 \sin 2 \beta_2 \\
{\cal I}_{13}^\tau  = c_1^2 c_3^2 \sin 2 \gamma_3 
\ea $ 
& $\ba
\tilde{m}_1^\mu = b_1^2 /M_1 \\
\tilde{m}_1^\tau = c_1^2 /M_1 
\ea $ 
&
$ \ba 
b_2^2  {M_1} {M_3} \sin 2 {\gamma_3}  -c_3^2 {M_1}
    {M_2} \sin 2 {\beta_2} \\ 
-\left(a_1^2+b_1^2+c_1^2\right) 
{M_2}{M_3} \sin 2 ({\beta_2}-{\gamma_3})
\ea $
\\ \hline
$
\left( 
\bad 
a_1 & 0 & 0 \\
b_1 e^{i \beta_1} & 0 & b_3 \\
c_1 & c_2 e^{i \gamma_2} & 0  
\ea 
\right)$
& 
$ \ba 
{\cal I}_{13}^\mu  = -b_1^2 b_3^2 \sin 2 \beta_1 \\
{\cal I}_{12}^\tau  = c_1^2 c_2^2 \sin 2 \gamma_2 
\ea $ 
& $\ba
\tilde{m}_1^\mu = b_1^2 /M_1 \\
\tilde{m}_1^\tau = c_1^2 /M_1 
\ea $ 
& 
$ \ba 
 b_3^2 {M_1} {M_2} \sin 2 {\gamma_2} +c_2^2 {M_1} 
    {M_3} \sin 2 {\beta_1} \\ 
+\left(a_1^2+b_1^2+c_1^2\right) 
{M_2} {M_3} \sin 2 ({\beta_1}+{\gamma_2})
\ea $
\\ \hline
\end{tabular}
\caption{\label{tab:2a}Category (iia): The Dirac mass matrix, 
the non-zero expressions relevant for leptogenesis and 
the corresponding wash-out factors, 
and the relevant part of the invariant 
for CP violation in neutrino oscillations. 
For this category at low energy the correlation 
$|(m_\nu)_{11} \, (m_\nu)_{23}| - |(m_\nu)_{12} \, (m_\nu)_{13}| = 
{\rm arg} \left\{ (m_\nu)_{11} \, (m_\nu)_{23} \, 
(m_\nu)_{12}^\ast \, (m_\nu)_{13}^\ast \right\} = 0$ applies.}

\end{sidewaystable}

\begin{sidewaystable}
\begin{tabular}{|c|c|c|c|} \hline
$m_D = $ & leptogenesis & wash-out & $J_{CP}\propto $\\ \hline \hline 
$
\left( 
\bad 
0 & a_2 e^{i \alpha_2} & a_3 \\
0 & 0 & b_3 \\
c_1 e^{i \gamma_1} & 0 & c_3 
\ea 
\right)$
& $ \ba {\cal I}_{13}^\tau  = - c_1^2 c_3^2 \sin 2 \gamma_1 
\ea$
&  $\tilde{m}_1^\tau = c_1^2 /M_1$ 
& 
$ \ba 
a_2^2 M_1 M_3 \sin 2 \gamma_1  - c_1^2 M_2  M_3 
\sin 2 \alpha_2 \\ 
-\left(a_3^2+b_3^2+c_3^2\right) M_1
    M_2 \sin 2 (\alpha_2- \gamma_1)
\ea $
 \\ \hline 
$
\left( 
\bad 
a_1 e^{i \alpha_1} & 0 & a_3 \\
0 & 0 & b_3 \\
0 & c_2 e^{i \gamma_2} & c_3 
\ea 
\right)$ 
& $ \ba {\cal I}_{13}^e  = - a_1^2 a_3^2 \sin 2 \alpha_1 
\ea $
&  $\tilde{m}_1^e = a_1^2 /M_1$ 
& 
$ \ba 
a_1^2 M_2 M_3 \sin 2 \gamma_2 - c_2^2 M_1 M_3 
\sin 2 \alpha_1  \\ 
-\left(a_3^2+b_3^2+c_3^2\right) M_1
    M_2 \sin 2 (\alpha_1-\gamma_2 )
\ea $
 \\ \hline 
$
\left( 
\bad 
0 & a_2 e^{i \alpha_2} & a_3 \\
0 & b_2 & 0 \\
c_1 e^{i \gamma_1} & c_2 & 0
\ea 
\right)$ 
& $ \ba {\cal I}_{12}^\tau  = - c_1^2 c_2^2 \sin 2 \gamma_1 
\ea $
&  $\tilde{m}_1^\tau = c_1^2 /M_1$ 
& 
$ \ba 
 a_3^2 M_1 M_2 \sin 2 \gamma_1 + c_1^2 M_2 M_3 
\sin 2 \alpha_2 \\  
+\left(a_2^2+b_2^2+c_2^2\right) M_1
    M_3 \sin 2 (\alpha_2+\gamma_1)
\ea $
 \\ \hline 
$
\left( 
\bad 
a_1 & a_2 e^{i \alpha_2} & 0 \\
0 & b_2 & 0 \\
0 & c_2 & c_3 e^{i \gamma_3} 
\ea 
\right)$ 
& $ \ba {\cal I}_{12}^e  = a_1^2 a_2^2 \sin 2 \alpha_2 
\ea $ 
&  $\tilde{m}_1^e = a_1^2 /M_1$ 
& 
$ \ba 
a_1^2  M_2 M_3 \sin 2 \gamma_3 +  
c_3^2 M_1 M_2 \sin 2 \alpha_2 \\
+ \left(a_2^2+b_2^2+c_2^2\right) M_1
    M_3 \sin 2 (\alpha_2+\gamma_3       )
\ea $
 \\ \hline
$
\left( 
\bad 
a_1 & a_2 e^{i \alpha_2} & 0 \\
b_1 & 0 & 0 \\
c_1 & 0 & c_3 e^{i \gamma_3} 
\ea 
\right)$ 
& 
$ \ba
{\cal I}_{12}^e  = a_1^2 a_2^2 \sin 2 \alpha_2 \\
{\cal I}_{13}^\tau  = c_1^2 c_3^2 \sin 2 \gamma_3  
\ea$ 
&  $\ba 
\tilde{m}_1^e = a_1^2 /M_1 \\
\tilde{m}_1^\tau = c_1^2 /M_1
\ea $ 
& 
$ \ba 
a_2^2 M_1 M_3 \sin 2 \gamma_3 - c_3^2 M_1 M_2 
\sin 2\alpha_2  \\ 
-\left(a_1^2+b_1^2+c_1^2\right) M_2
    M_3 \sin 2 (\alpha_2-\gamma_3)
\ea $
 \\ \hline 
$
\left( 
\bad 
a_1 e^{i \alpha_1} & 0 & a_3 \\
b_1 & 0 & 0 \\
c_1 & c_2 e^{i \gamma_2} & 0
\ea 
\right)$ 
& 
$ \ba
{\cal I}_{13}^e  = -a_1^2 a_3^2 \sin 2 \alpha_1 \\
{\cal I}_{12}^\tau  = c_1^2 c_2^2 \sin 2 \gamma_2 
\ea$ 
&  $\ba 
\tilde{m}_1^e = a_1^2 /M_1 \\
\tilde{m}_1^\tau = c_1^2 /M_1
\ea $ 
& 
$ \ba 
a_3^2 M_1 M_2 \sin 2 \gamma_2  + c_2^2 M_1  M_3 
\sin 2 \alpha_1 \\ + \left(a_1^2+b_1^2+c_1^2\right) M_2
    M_3 \sin 2 (\alpha_1 + \gamma_2)
\ea $
 \\ \hline 
\end{tabular}
\caption{\label{tab:2b}Category (iib): the Dirac mass matrix, 
the non-zero expressions 
relevant for leptogenesis and the corresponding wash-out factors, 
and the relevant part of the invariant 
for CP violation in neutrino oscillations. 
For this category at low energy the correlation 
$|(m_\nu)_{22} \, (m_\nu)_{13}| - |(m_\nu)_{12} \, (m_\nu)_{23}| = 
{\rm arg} \left\{ (m_\nu)_{22} \, (m_\nu)_{13} 
(m_\nu)_{12}^\ast \, (m_\nu)_{23}^\ast \right\} = 0$ applies.}

\end{sidewaystable}

\begin{sidewaystable}
\begin{tabular}{|c|c|c|c|} \hline
$m_D = $ & leptogenesis & wash-out & $J_{CP} \propto $\\ \hline \hline 
$
\left( 
\bad 
a_1 e^{i \alpha_1} & 0 & a_3 \\
0 & b_2 e^{i \beta_2} & b_3 \\
0 & 0 & c_3 
\ea 
\right)$ 
& $\ba {\cal I}_{13}^e  = - a_1^2 a_3^2 \sin 2 \alpha_1 
\ea $
&  $\tilde{m}_1^e = a_1^2 /M_1$ 
& 
$ \ba 
a_1^2  M_2 M_3 \sin 2 \beta_2  
-b_2^2 M_1  M_3 \sin 2 \alpha_1 \\
-\left(a_3^2+b_3^2+c_3^2\right) M_1
    M_2 \sin 2 (\alpha_1-\beta_2)

\ea $
 \\ \hline 
$
\left( 
\bad 
0 & a_2 e^{i \alpha_2} &  a_3 \\
b_1 e^{i \beta_1} & 0 & b_3 \\
0 & 0 & c_3 
\ea 
\right)$ 
& $ \ba {\cal I}_{13}^\mu  = - b_1^2 b_3^2 \sin 2 \beta_1 
\ea $
&  $\tilde{m}_1^\mu = b_1^2 /M_1$ 
& 
$ \ba 
a_2^2  M_1 M_3 \sin 2 \beta_1 
-b_1^2 M_2  M_3 \sin 2\alpha_2  \\ 
-\left(a_3^2+b_3^2+c_3^2\right) M_1 M_2 
\sin 2 (\alpha_2-\beta_1 )
\ea $ \\ \hline
$
\left( 
\bad 
a_1 e^{i \alpha_1} & a_2 & 0 \\
0 & b_2 e^{i \beta_2} & b_3 \\
0 & c_2 & 0 
\ea 
\right)$ 
& $ \ba {\cal I}_{12}^e  = - a_1^2 a_2^2 \sin 2 \alpha_1  
\ea $
&  $\tilde{m}_1^e = a_1^2 /M_1$ 
& 
$ \ba 
 a_1^2 M_2 M_3 \sin 2 \beta_2  
+b_3^2 M_1
    M_2 \sin 2 \alpha_1  \\
+\left(a_2^2+b_2^2+c_2^2\right) M_1 M_3 
\sin 2 (\alpha_1+\beta_2)
\ea $ \\ \hline
$
\left( 
\bad 
0 & a_2  & a_3 e^{i \alpha_3}\\
b_1 & b_2 e^{i \beta_2} & 0 \\
0 & c_2 & 0 
\ea 
\right)$ 
& ${\cal I}_{12}^\mu  = b_1^2 b_2^2 \sin 2 \beta_2$
&  $\tilde{m}_1^\mu = b_1^2 /M_1$ 
& 
$ \ba 
a_3^2 M_1 M_2 \sin 2 \beta_2 
+b_1^2 M_2
    M_3 \sin 2\alpha_3 \\ 
+\left(a_2^2+b_2^2+c_2^2\right) M_1
    M_3 \sin 2 (\alpha_3+\beta_2)
\ea $ \\ \hline
$
\left( 
\bad 
a_1 & 0 & a_3 e^{i \alpha_3} \\
b_1 & b_2 e^{i \beta_2} &  0 \\
c_1 & 0 & 0 
\ea 
\right)$ 
& $\ba 
{\cal I}_{13}^e  = a_1^2 a_3^2 \sin 2 \alpha_3 \\
{\cal I}_{12}^\mu  = b_1^2 b_2^2 \sin 2 \beta_2 
\ea $
&  $\ba 
\tilde{m}_1^e = a_1^2 /M_1 \\
\tilde{m}_1^\mu = b_1^1 /M_1 
\ea $ 
& 
$ \ba 
a_3^2 M_1 M_2 \sin 2 \beta_2 
-b_2^2 M_1
    M_3 \sin 2 \alpha_3 \\
-\left(a_1^2+b_1^2+c_1^2\right) M_2
    M_3 \sin 2 (\alpha_3-\beta_2)
\ea $ \\ \hline
$
\left( 
\bad 
a_1 & a_2 e^{i \alpha_2} & 0\\
b_1 e^{i \beta_1} &  0 & b_3 \\
c_1 & 0 & 0 
\ea 
\right)$ 
& $\ba 
{\cal I}_{12}^e  = a_1^2 a_2^2 \sin 2 \alpha_2 \\
{\cal I}_{13}^\mu  = -b_1^2 b_3^2 \sin 2 \beta_1 
\ea $
&  $\ba 
\tilde{m}_1^e = a_1^2 /M_1 \\
\tilde{m}_1^\mu = b_1^1 /M_1 
\ea $ 
& 
$ \ba 
a_2^2 M_1 M_3 \sin 2 \beta_1 
+b_3^2 M_1
    M_2 \sin 2 \alpha_2 \\
+\left(a_1^2+b_1^2+c_1^2\right) M_2
    M_3 \sin 2 (\alpha_2+\beta_1)
\ea $ \\ \hline
\end{tabular}
\caption{\label{tab:2c}Category (iic): the Dirac mass matrix, 
the non-zero expressions 
relevant for leptogenesis and the corresponding wash-out factors, 
and the relevant part of the invariant 
for CP violation in neutrino oscillations. 
For this category at low energy the correlation 
$|(m_\nu)_{33} \, (m_\nu)_{12}| - |(m_\nu)_{13} \, (m_\nu)_{32}| = 
{\rm arg} \left\{ (m_\nu)_{33} \, (m_\nu)_{12} 
(m_\nu)_{13}^\ast \, (m_\nu)_{32}^\ast \right\} = 0$ applies.}

\end{sidewaystable}

\end{document}